\documentstyle[epsf,iopconf1]{article}
\newcommand{\be}[1]{\begin{equation} \label{(#1)}}
\newcommand{\ee}{\end{equation}}
\newcommand{\ba}[1]{\begin{eqnarray} \label{(#1)}}
\newcommand{\ea}{\end{eqnarray}}
\newcommand{\nn}{\nonumber}

\def \znbb {$0\nu\beta\beta$}

\def \Rpv{R_{P} \hspace{-0.9em}/\;\:}
\def\rp{$R_p \hspace{-1em}/\;\:$}
\def \emass {\langle m_{\nu} \rangle}

\def\be{\begin{equation}}
\def\ee{\end{equation}}
\def\bea{\begin{eqnarray}}
\def\eea{\end{eqnarray}}

\begin{document}

\title{Perspectives of Double Beta and Dark Matter Search as Windows to 
New Physics}

\author{
H.V. Klapdor--Kleingrothaus}

\affil{\dag\ {Max--Planck--Institut f\"ur Kernphysik\\
P.O.Box 10 39 80, D--69029 Heidelberg, Germany} \\
}

\beginabstract
Nuclear double beta decay provides an extraordinarily broad potential
to search for beyond Standard Model physics, probing already now the
TeV scale, on which new physics should manifest itself. These possibilities
are reviewed here.
First, the results of present generation experiments are presented.
The most sensitive one of them -- the Heidelberg--Moscow experiment in the 
Gran Sasso -- probes the electron mass now in the sub eV region and 
has reached recently a limit of $\sim$ 0.1 eV.
This limit has striking influence on presently discussed neutrino mass 
scenarios. 
Basing to a large extent on the theoretical
work of the Heidelberg Double Beta Group in the last two years, results
are obtained also for SUSY models (R--parity breaking, sneutrino mass),
leptoquarks (leptoquark-Higgs coupling), compositeness,
right--handed W boson mass, test of special relativity and equivalence 
principle in the neutrino sector and others. These results are comfortably 
competitive to corresponding results from high--energy accelerators like TEVATRON, HERA, etc. One of the enriched $^{76}$Ge detectors also yields the
most stringent limits for cold dark matter (WIMPs) to date by using raw data.
Second, future perspectives of $\beta\beta$ research are discussed.  A new
Heidelberg experimental proposal (GENIUS) will allow to 
increase the sensitivity for Majorana neutrino masses from the present level
of at best 0.1 eV down to 0.01 or even 0.001 eV. Its physical potential would 
be a breakthrough into the multi-TeV range for many beyond standard models.
Its sensitivity for neutrino oscillation parameters would be larger than of 
all present terrestrial neutrino oscillation experiments and of those planned 
for the future. It could probe directly the large angle, and for almost 
degenerate neutrino mass scenarios even the small angle
solution of the solar neutrino problem. It would 
further, already in a first step using only 100 kg of natural Ge detectors, 
cover almost the full MSSM parameter 
space for 
prediction of neutralinos as cold dark matter, making the experiment 
competitive to LHC in the search for supersymmetry. Finally GENIUS could be used as the first real time detector of solar pp neutrinos. 

\endabstract

\section{Introduction -- Motivation for the search for double beta decay
-- and a future perspective: GENIUS}
Double beta decay yields -- besides proton decay -- the most promising
possibilities to probe beyond standard model physics beyond accelerator 
energy scales. 

The potential of double beta decay includes information on the neutrino and
sneutrino mass, SUSY models, 
compositeness, leptoquarks, right--handed $W$ bosons, Lorentz invariance and 
the equivalence principle in the neutrino sector, and others \cite{Kla99a}. 
The recent results of the Heidelberg--Moscow experiment, 
which will be reported here (see also \cite{KK1,KK2}),
have demonstrated 
that $0\nu\beta\beta$ decay probes already now the TeV
scale on which new physics should manifest itself according to present 
theoretical expectations.

To increase by a major step the 
present sensitivity for double beta decay and dark matter search, we describe
here a new project proposed recently \cite{KK1,KK2}
which would operate one ton
of `naked' enriched {\bf GE}rmanium detectors in liquid 
{\bf NI}trogen as shielding in an {\bf U}nderground {\bf S}etup (GENIUS).
GENIUS would definitely be a breakthrough into the multi-TeV
range for many beyond standard models currently discussed in the literature,
and the sensitivity would be comparable or even superior to LHC for various
quantities such as right--handed W--bosons, R--parity violation, leptoquark
or compositeness searches.

Another issue of GENIUS is the search for Dark Matter in the universe.
The full
MSSM parameter space for predictions of neutralinos as cold 
dark matter could be 
covered already in a first step of the full experiment using only 100 kg
of $^{76}$Ge or even natural Ge, making the experiment competitive to LHC
in the search for supersymmetry. 

Finally GENIUS could be used as the first real time detector of solar pp neutrinos.

\section{Double beta decay and particle physics}
We present a brief introductory outline of the potential of $\beta\beta$ decay
for some representative examples.
The potential of double beta decay for probing neutrino oscillation
parameters will be addressed in section 4.2.

Double beta decay can occur in several decay modes
\be
^{A}_{Z}X \rightarrow ^A_{Z+2}X + 2 e^- + 2 {\overline \nu_e}
\ee
\be        
^{A}_{Z}X \rightarrow ^A_{Z+2}X + 2 e^- 
\ee
\be
^{A}_{Z}X \rightarrow ^A_{Z+2}X + 2 e^- + \phi
\ee
\be
^{A}_{Z}X \rightarrow ^A_{Z+2}X + 2 e^- + 2\phi
\ee
the last three of them violating lepton number conservation by $\Delta L=2$.
For the neutrinoless mode
(2) we expect a sharp line at $E=Q_{\beta\beta}$, for the two--neutrino mode
and the various Majoron--accompanied modes classified by their spectral index,
continuous spectra.
Important for particle physics are the decay modes (2)--(4).

The neutrinoless mode (2) needs not be necessarily connected with the 
exchange of a virtual neutrino or sneutrino. {\it Any} process violating 
lepton number can
in principle lead to a process with the same signature as usual 
$0\nu\beta\beta$
decay. It may be triggered by exchange of neutralinos, gluinos, squarks,
sleptons, leptoquarks,... (see below and \cite{KK2,Paes97,Paes99}). 
This gives rise
to the broad potential of double beta decay for testing or yielding 
restrictions on
quantities of beyond standard model physics, realized and 
investigated to a large extent by the Heidelberg Double Beta Group in the last 
two years. 
There is, however, a generic relation between the amplitude of $0\nu\beta\beta$
decay and the $(B-L)$ violating Majorana mass of the neutrino. It has been 
recognized about 15 years ago \cite{Sch81} that if any of these two quantities
vanishes, the other one vanishes, too, and vice versa, if one of them is
non--zero, the other one also differs from zero. This Schechter-Valle-theorem 
is valid for
any gauge model with spontaneously broken symmetry at the weak scale,
independent of the mechanism of $0\nu\beta\beta$ decay. A generalisation
of this theorem to supersymmetry has been given recently \cite{Hir97,Hir97a}.
This Hirsch--Klapdor-Kleingrothaus--Kovalenko--theorem claims for the neutrino 
Majorana mass, the $B-L$ violating mass of the
sneutrino and neutrinoless double beta decay amplitude:
If one of them is non--zero, also the others are non--zero and vice versa,
independent of the mechanisms of $0\nu\beta\beta$ decay and (s-)neutrino
mass generation. This theorem connects double beta research with new processes
potentially observable at future colliders like NLC (next linear collider)
\cite{Hir97,Kolb1}.


\subsection{Mass of the (electron) neutrino}
Neutrino physics has entered an era of new actuality in connection
with several possible indications of physics beyond the standard model
(SM) of particle physics: 
A lack of solar
($^7Be$) neutrinos, an atmospheric $\nu_{\mu}$ deficit and mixed dark matter 
models could all be explained simultaneously by non--vanishing neutrino masses.
Recent GUT models, for example an extended SO(10) scenario with $S_4$
horizontal symmetry could explain these observations by requiring 
degenerate neutrino masses of the order of 1 eV 
\cite{19,Moh94,20,21,22,23} \cite{12,13}.
For an overview see \cite{Smi96a,Mohneu}.

This brings double beta decay experiments into some key
position, since with some second generation $\beta\beta$ experiments like the 
HEIDELBERG--MOS\-COW experiment the predictions of or assumptions in such
scenarios can now be tested (see section 3.2). 
If the above scenario of neutrino
mass textures is ruled out by tightening the double beta limit on $m_{\nu_e}$,
then a way to understand {\it all} neutrino results may 
require an 
additional sterile neutrino \cite{Cal93,Pel93,Mohneu}. 
Then the solar neutrino puzzle could be explained by the
$\nu_e-\nu_S$ oscillation, and atmospheric neutrino data by 
$\nu_{\mu}-\nu_{\tau}$ oscillations, and the $\nu_{\mu,\tau}$ would constitute
the hot dark matter (HDM) of the universe. The request for a light sterile
neutrino would naturally lead to the concept of a shadow world \cite{Ber95}.
The expectation for the effective
neutrino mass (see below) to be seen in double beta decay would be 
$ \langle m_{\nu_{e}} \rangle \simeq 0.002 eV$ \cite{Moh97a}. 
Thus it could
be checked by the new Genius project (see section 4.2.2). 

Neutrinoless double beta decay can be triggered by exchange of a
light or heavy left-handed Majorana neutrino.
For exchange of a heavy {\it right}--handed neutrino see section 2.3.
      The propagators in the first and second case show a different $m_{\nu}$
dependence: Fermion propagator $\sim \frac {m}{q^2-m^2} \Rightarrow$
\be
a)\hskip5mm m\ll q \rightarrow \sim m \hskip5mm 'light' \hskip2mm neutrino
\ee
\be
b)\hskip5mm m\gg q \rightarrow \sim \frac{1}{m} \hskip5mm 'heavy' \hskip2mm
neutrino
\ee   
The half--life for $0\nu\beta\beta$ decay induced by exchange of a light 
neutrino is given by \cite{27}
\ba{71}
[T^{0\nu}_{1/2}(0^+_i \rightarrow 0^+_f)]^{-1}= C_{mm} 
\frac{\langle m_{\nu} \rangle^2}{m_{e}^2}
+C_{\eta\eta} \langle \eta \rangle^2 + C_{\lambda\lambda} 
\langle \lambda \rangle^2 +C_{m\eta} \frac{m_{\nu}}{m_e} 
\nn
\ea
\be
+ C_{m\lambda}
\langle \lambda \rangle \frac{\langle m_{\nu} \rangle}{m_e} 
+C_{\eta\lambda} 
\langle \eta \rangle \langle \lambda \rangle
\ee
or, when neglecting the effect of right--handed weak currents, by
\be
[T^{0\nu}_{1/2}(0^+_i \rightarrow 0^+_f)]^{-1}=C_{mm} 
\frac{\langle m_{\nu} \rangle^2}{m_{e}^2}
=(M^{0\nu}_{GT}-M^{0\nu}_{F})^2 G_1 
\frac{\langle m_{\nu} \rangle^2}{m_e^2}
\ee
where $G_1$ denotes the phase space integral, $ \langle m_{\nu} \rangle$
denotes an effective neutrino mass
\be
\langle m_{\nu} \rangle = \sum_i m_i U_{ei}^2,
\ee
respecting the possibility of the electron neutrino to be a mixed state
(mass matrix not diagonal in the flavor space)
\be
|\nu_e \rangle = \sum_i U_{ei} |\nu_{i}\rangle
\ee

The effective mass $\langle m_{\nu} \rangle$ could be smaller than $m_i$
for all i for appropriate CP phases of the mixing coefficiants $U_{ei}$
\cite{Wol81}.
In general not too pathological GUT models yield 
$m_{\nu_e}=\langle m_{\nu_e}
\rangle$ (see \cite{15}).

$\eta$,$\lambda$ describe an admixture of right--handed weak currents, and
$M^{0\nu}\equiv M_{GT}^{0\nu}-M_{F}^{0\nu}$ denote nuclear matrix elements.

\subsection*{Nuclear matrix elements:}
A detailed discussion of $\beta\beta$ matrix elements for neutrino induced
transitions including the substantial (well--understood) differences
in the precision with which $2\nu$ and $0\nu\beta\beta$ rates can be 
calculated, can be found in \cite{16,27,28} \cite{29,KK1,KK2}.

\subsection{Supersymmetry}
Supersymmetry (SUSY) is considered as prime candidate for a theory beyond the
standard model, which could overcome some of the most puzzling questions of
today's particle physics (see, e.g. \cite{44,45,Kan97}). 
Generally one can add the following R--parity violating terms  
to the usual superpotential \cite{hal84}.
\be
W_{\Rpv}=\lambda_{ijk}L_{i}L_{j}\overline{E}_{k}+\lambda^{'}_{ijk}
L_i Q_j \overline{D}_k + \lambda^{''}\overline{U}_i \overline{D}_j 
\overline{D}_k,
\ee
where indices $i,j,k$ denote generations. $L$,$Q$ denote lepton and quark 
doublet superfields and $\overline{E}, \overline{U}, \overline{D}$ lepton and
up, down quark singlet superfields. Terms proportional to $\lambda$, 
$\lambda^{'}$
violate lepton number, those proportional to $\lambda^{''}$ violate baryon 
number. From proton decay limits it is clear that both types of terms cannot 
be present at the same time in the superpotential. On the other hand, once the
$\lambda^{''}$ terms being assumed to be zero, $\lambda$ and $\lambda^{'}$
terms are not limited. $0\nu\beta\beta$ decay can occur within the 
\rp MSSM through Feynman graphs  such as those of Fig. 1. 
In lowest order 
there are alltogether six different graphs of this kind. \cite{6,47,75}.    
Thus  $0\nu\beta\beta$ decay can be used to restrict R--parity violating
SUSY models \cite{6,hir96c,17,47,48}. From these graphs one derives \cite{6}
under some assumptions 
\be
[T^{0\nu}_{1/2}(0^+ \rightarrow 0^+)]^{-1} \sim G_{01}
(\frac{\lambda_{111}'^2}{m^4_{{\tilde q},{\tilde e}}m_{{\tilde g}\chi}}M)^2
\ee
where $G_{01}$ is a phase space factor, 
$m_{{\tilde q}{\tilde e}{\tilde g}\chi}$
are the masses of supersymmetric particles involved: squarks, selectrons,
gluinos, or neutralinos. $\lambda'_{111}$ is the strength of an R--parity
breaking interaction (eq. 11), and $M$ is a nuclear matrix element. For the
matrix elements and their calculation see \cite{hir96c}.

\begin{figure}
\parbox{14cm}{
\epsfxsize=50mm
\epsfbox{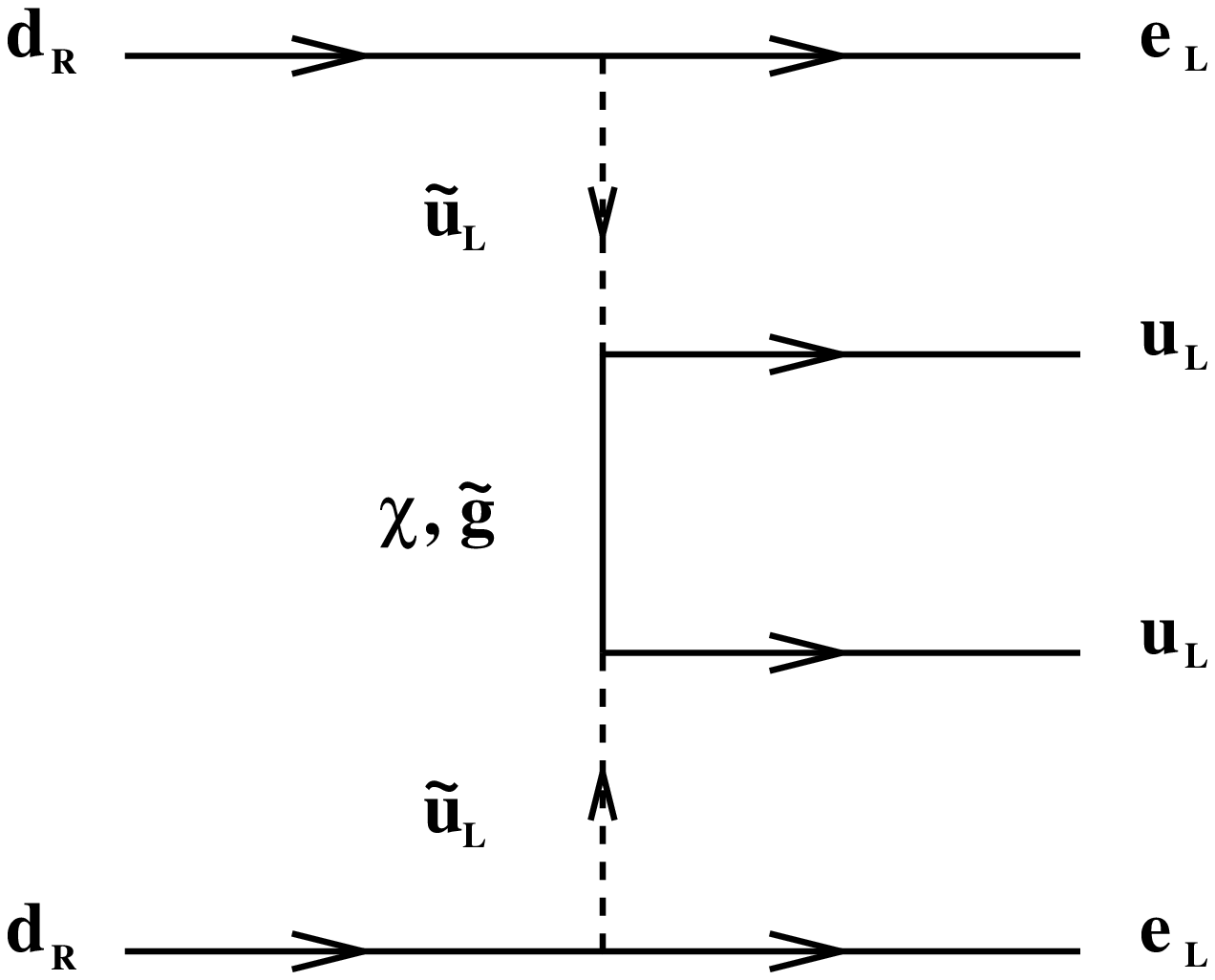}
\parbox{6cm}{
\vspace*{-45mm}
\hspace*{60mm}
\epsfxsize=50mm
\epsfbox{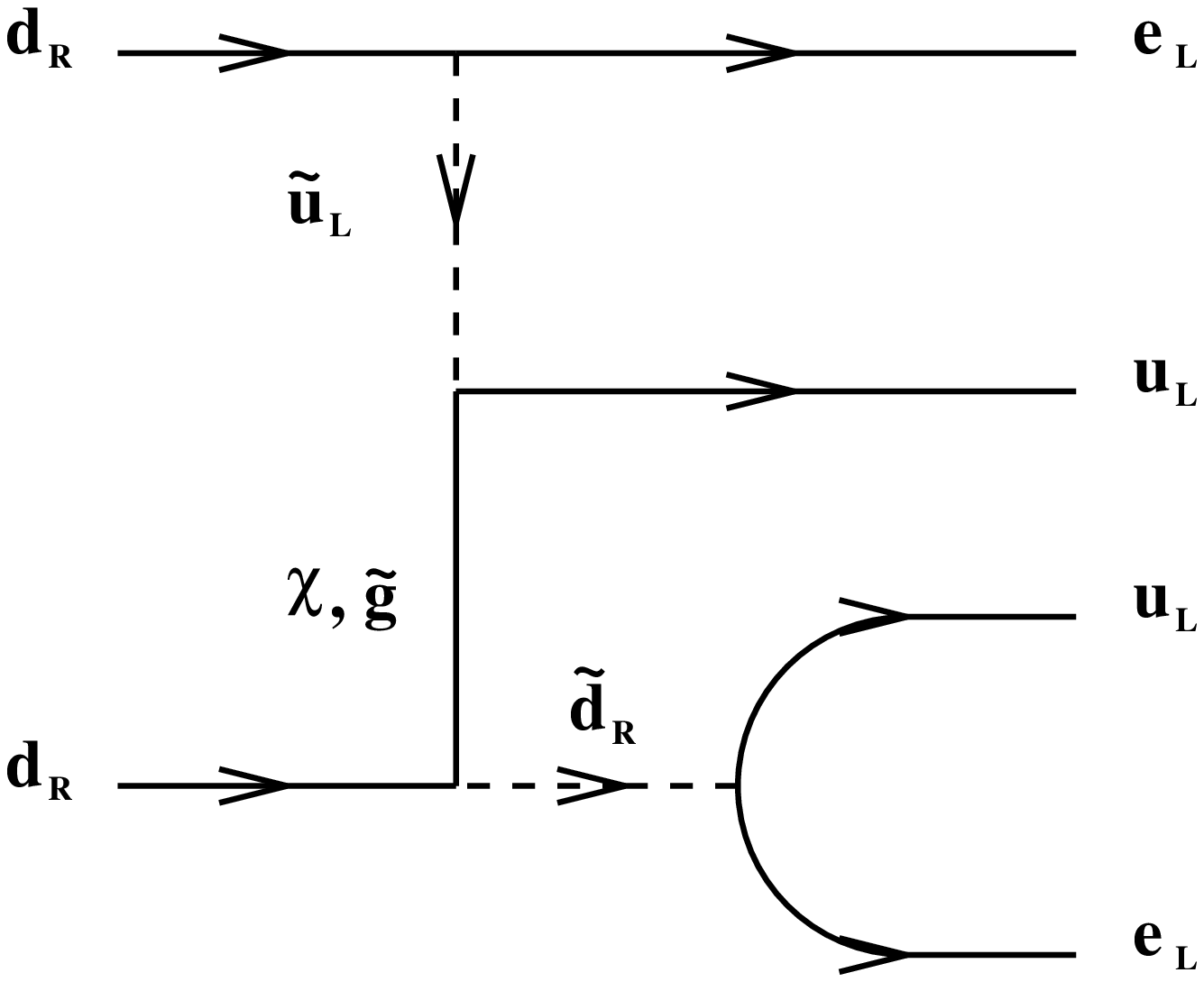}
}

{\bf Fig. 1} {\it Examples of Feynman graphs for $0\nu\beta\beta$ 
decay within R--parity \\
violating supersymmetric models (from [Hir95a]).}}



\parbox{14cm}{
\vspace*{6mm}
\epsfxsize=50mm
\epsfbox{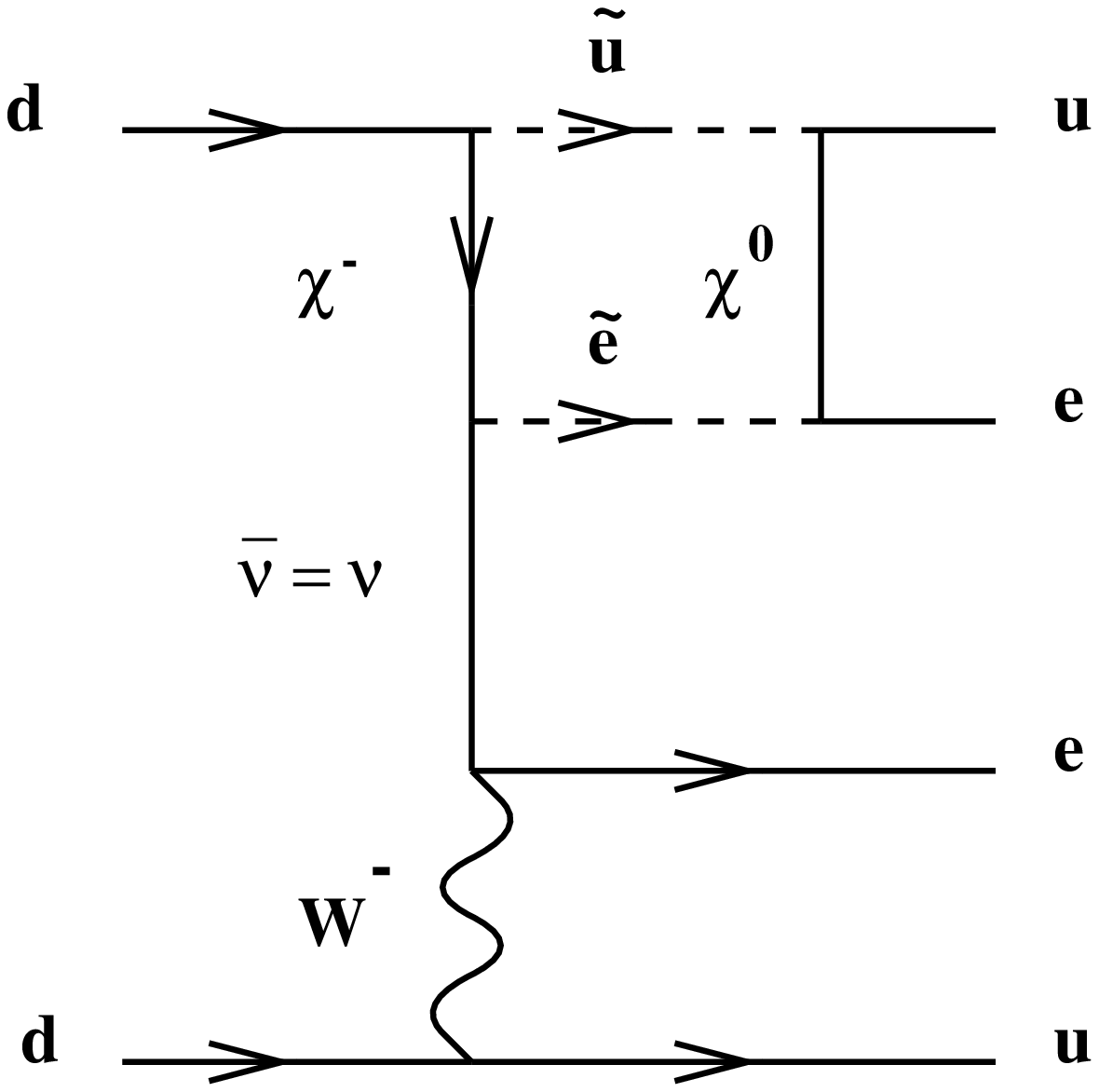}
\parbox{6cm}{
\vspace*{-45mm}
\hspace*{60mm}
\epsfxsize=50mm
\epsfbox{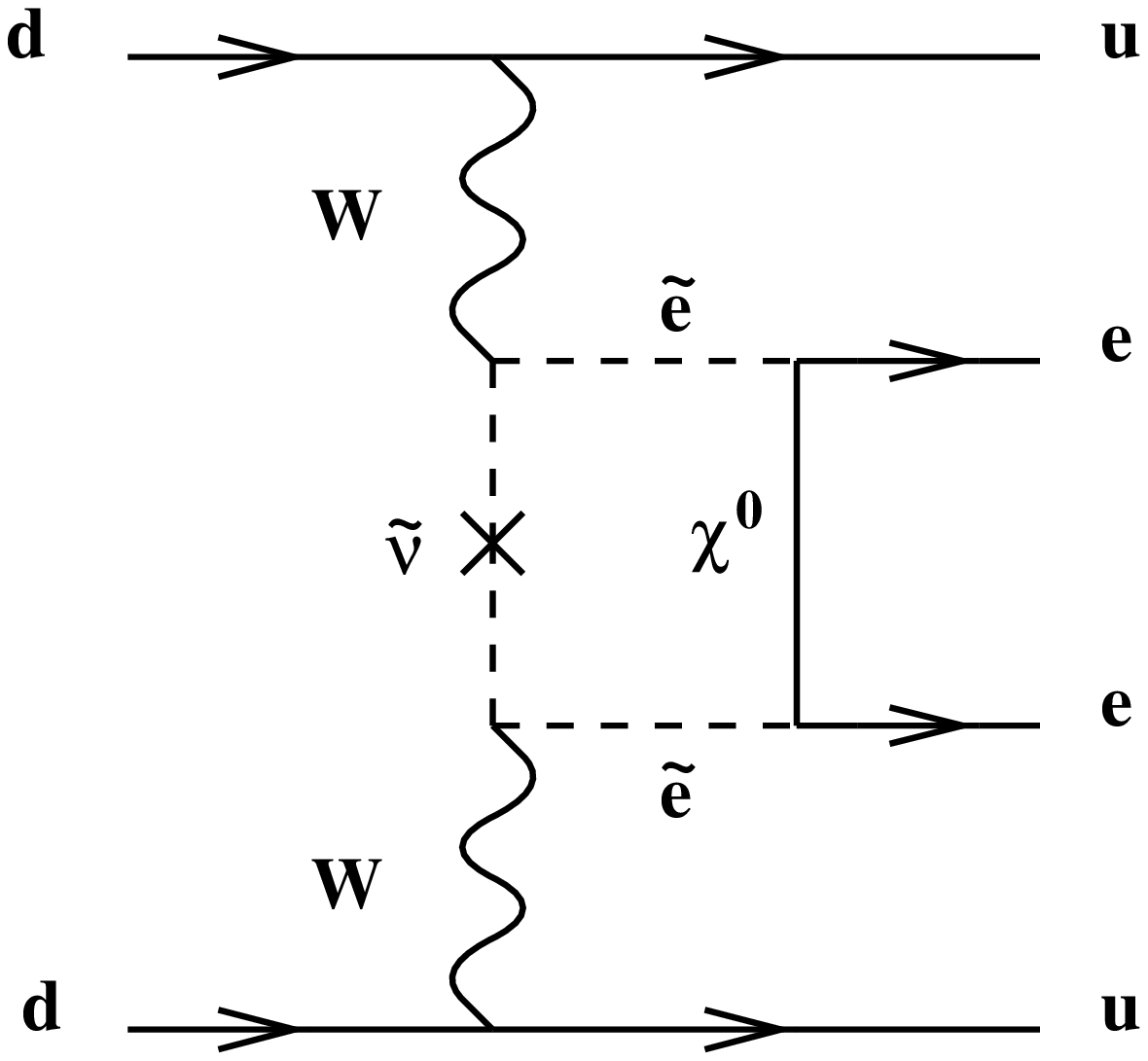}
\vspace*{8mm}
}

{\bf Fig. 2} {\it Examples of $R_P$ conserving SUSY contributions
to $0\nu\beta\beta$ decay \\
(from [Hir97a]).}}   
\end{figure}

It is also worthwile to notice that $0\nu\beta\beta$ decay is not only 
sensitive to $\lambda^{'}_{111}$. Taking into account the fact that the SUSY
partners of the left and right--handed quark states can mix with each other, 
one can derive limits on different combinations of $\lambda^{'}$ 
\cite{hir96,7,bab95}.
The dominant diagram 
of this type is the one where the exchanged scalar particles are the
$\tilde{b}-\tilde{b}^C$ pair. Under some assumptions (e.g. the MSSM mass 
parameters to be approximately equal to the ``effective'' SUSY breaking scale 
$\Lambda_{SUSY}$), one obtains \cite{hir96}
\be
\lambda_{11i}^{'}\cdot \lambda_{1i1}^{'}\leq \epsilon_i^{'} 
\Big( \frac{\Lambda_{SUSY}}{100 GeV} \Big)^3
\ee
and
\be
\Delta_n \lambda^{'}_{311} \lambda_{n13} \leq \epsilon \Big(\frac
{\Lambda_{SUSY}
}{100 GeV}\Big)^3
\ee
For an overview on our knowledge
on $\lambda^{'}_{ijk}$
from other sources we refer to \cite{Kol97a} and \cite{Bha97}.

Also R--parity {\it conserving} softly broken supersymmetry can give 
contributions to $0\nu\beta\beta$ decay, via the $B-L$--violating sneutrino 
mass term, the latter being a generic ingredient of any weak--scale SUSY
model with a Majorana neutrino mass \cite{Hir97,Kolb1}. 
These contributions are 
realized  at the level of box diagrams \cite{Kolb1} (fig. 2).
The $0\nu\beta\beta$ half-life for contributions from sneutrino exchange
is found to be \cite{Kolb1}
\be
[{T_{1/2}^{0\nu\beta\beta}}]^{-1}=G_{01}\frac{4 m_p^2}{G^4_F} 
\Big|\frac{\eta^{SUSY}}{m^5_{SUSY}} M^{SUSY}\Big|,
\ee
where the phase factor $G_{01}$ is tabulated in \cite{74}, $\eta^{SUSY}$
is the effective lepton number violating parameter, which contains the
$(B-L)$ violating sneutrino mass $\tilde{m}_M$ and $M^{SUSY}$ is the nuclear 
matrix element \cite{11}.



\subsection{Left--Right symmetric theories --
Heavy neutrinos and right--handed W Boson}
Heavy {\it right--handed } neutrinos appear quite naturally in left--right
symmetric GUT models. Since in such models the symmetry breaking
scale for the right--handed sector is not fixed by the theory, the
mass of the right--handed $W_R$ boson and the mixing angle between the mass 
eigenstates $W_1$, $W_2$ are free parameters. $0\nu\beta\beta$ decay taking
into account contributions from both, left-- and right--handed neutrinos
have been studied theoretically by \cite{11,49}. The former gives a more
general expression for the decay rate than introduced earlier by \cite{50}. 

The amplitude will be proportional to \cite{11}
\be
\Big( \frac{m_{W_{L}}}{m_{W_R}}^4 \Big) \Big(\frac{1}{m_N}+\frac{m_N}
{m^2_{\Delta^{--}_R}}\Big)
\label{ncs3}
\ee

Eq. \ref{ncs3} and the experimental lower limit of $0\nu\beta\beta$ decay leads 
to a constraint limit within the 3--dimensional parameter space
($m_{W_R}-m_N-m_{\Delta^{--}_R}$). 


\subsection{Compositeness}
Although so far there are no experimental signals of a substructure of quarks 
and leptons, there are speculations that at some higher energy ranges beyond 1 
TeV or so there might exist an energy scale $\Lambda_C$ at which a
substructure of quarks and leptons (preons) might become visible 
\cite{8,45,51,Pan99}.

A possible low energy manifestation of compositeness could be neutrinoless
double beta decay, mediated by a composite heavy Majorana neutrino, 
which then should be a Majorana particle.

Recent theoretical work shows (see \cite{8,9,Pan97,Tak97,Pan99}) 
that the mass bounds for such an excited neutrino 
which can be derived from double 
beta decay are at
 least of the same order of magnitude as those coming from the
direct search of excited states in high energy accelerators 
(see also section 3).

\subsection{Majorons} 
The existence of new bosons, so--called Majorons, can play a significant
role in new physics beyond the standard model, in the history
of the early universe, in the evolution of stellar objects, in supernovae
astrophysics and the solar neutrino problem \cite{61,62,Kla92}.
In many theories of physics beyond the standard model neutrinoless 
double beta decay can occur with the emission of Majorons  
\be
2n\rightarrow2p+2e^{-}+\phi
\ee
\be
2n\rightarrow2p+2e^{-}+2\phi.
\ee 

To avoid an unnatural fine--tuning in recent years several 
new Majoron models were proposed \cite{68,69,70}, 
where the term
Majoron denotes in a more general sense light or massless bosons 
with couplings to neutrinos. 

The main novel features of these ``New Majorons'' are that they
can carry leptonic charge, that they need not be 
Goldstone bosons and that emission of two Majorons 
can occur. 
The latter can be scalar--mediated 
or fermion--mediated. For details we refer to
\cite{71,72}. 

The half--lifes are according to \cite{73,74} in some approximation given
by  
\be
[T_{1/2}]^{-1}=|<g_{\alpha}>|^{2}\cdot|M_{\alpha}|^{2}\cdot G_{BB_{\alpha}}
\ee
for $\beta\beta\phi$-decays, or
\be
[T_{1/2}]^{-1}=|<g_{\alpha}>|^{4}\cdot|M_{\alpha}|^{2}\cdot G_{BB_{\alpha}}
\ee
for $\beta\beta\phi\phi$--decays. The index ${\alpha}$ 
indicates that effective neutrino--Majoron coupling constants $g$, 
matrix elements $M$ and phase spaces $G$ differ for different models.

\subsection*{Nuclear matrix elements:}
There are five different nuclear matrix elements. Of
 these $M_{F}$ and $M_{GT}$ are the same which occur in $0\nu\beta\beta$ decay.
The other ones and the corresponding phase spaces have been calculated 
for the first time
by \cite{71,75}. The calculations of the matrix elements show 
that the new models predict, 
as consequence of the small matrix elements 
 very large half--lives and that unlikely large
coupling constants would be needed to produce observable decay rates
(see \cite{71,75}).

\subsection{Sterile neutrinos}

Introduction of sterile neutrinos has been claimed to solve simultaneously the 
conflict between dark matter neutrinos, LSND and supernova nucleosynthesis
\cite{76} and light sterile neutrinos are part of popular 
neutrino mass textures
for understanding the various hints for neutrino
oscillations (see section 2.1) and \cite{Moh96,Mohneu,Moh97a}. 
Neutrinoless double beta decay can also
investigate several effects
of {\it heavy} sterile neutrinos \cite{77}.

If we assume having a light neutrino with a mass $\ll$ 1 eV, mixing with a much
 heavier (m $\ge$ 1 GeV) sterile neutrino can yield under certain conditions
a detectable signal in current $\beta\beta$ experiments.


\subsection{Leptoquarks}

Interest on leptoquarks (LQ) has been renewed during the last few years 
since ongoing collider experiments have good prospects for searching 
these particles \cite{Lagr1}. LQs are vector or scalar particles 
carrying both lepton and baryon numbers and, therefore, have a 
well distinguished experimental signature. Direct searches of LQs in 
deep inelastic ep-scattering at HERA \cite{H196} placed lower limits 
on their mass $M_{LQ} \ge 225-275$ GeV, depending on the LQ type and 
couplings. 

\begin{figure}
\parbox{7cm}{
\epsfxsize=50mm
\epsfysize=50mm
\epsfbox{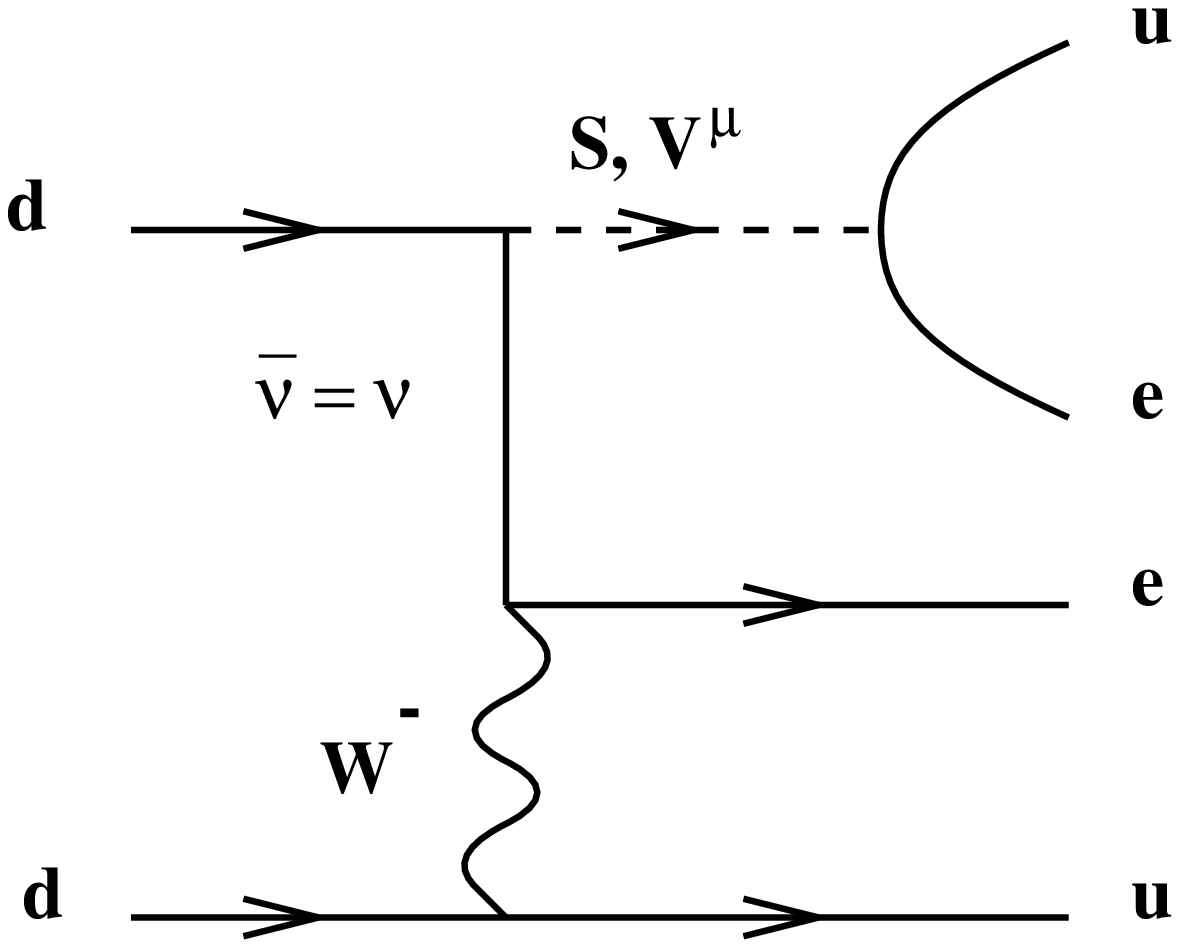}}
\parbox{7cm}{
\epsfxsize=50mm
\epsfysize=50mm
\epsfbox{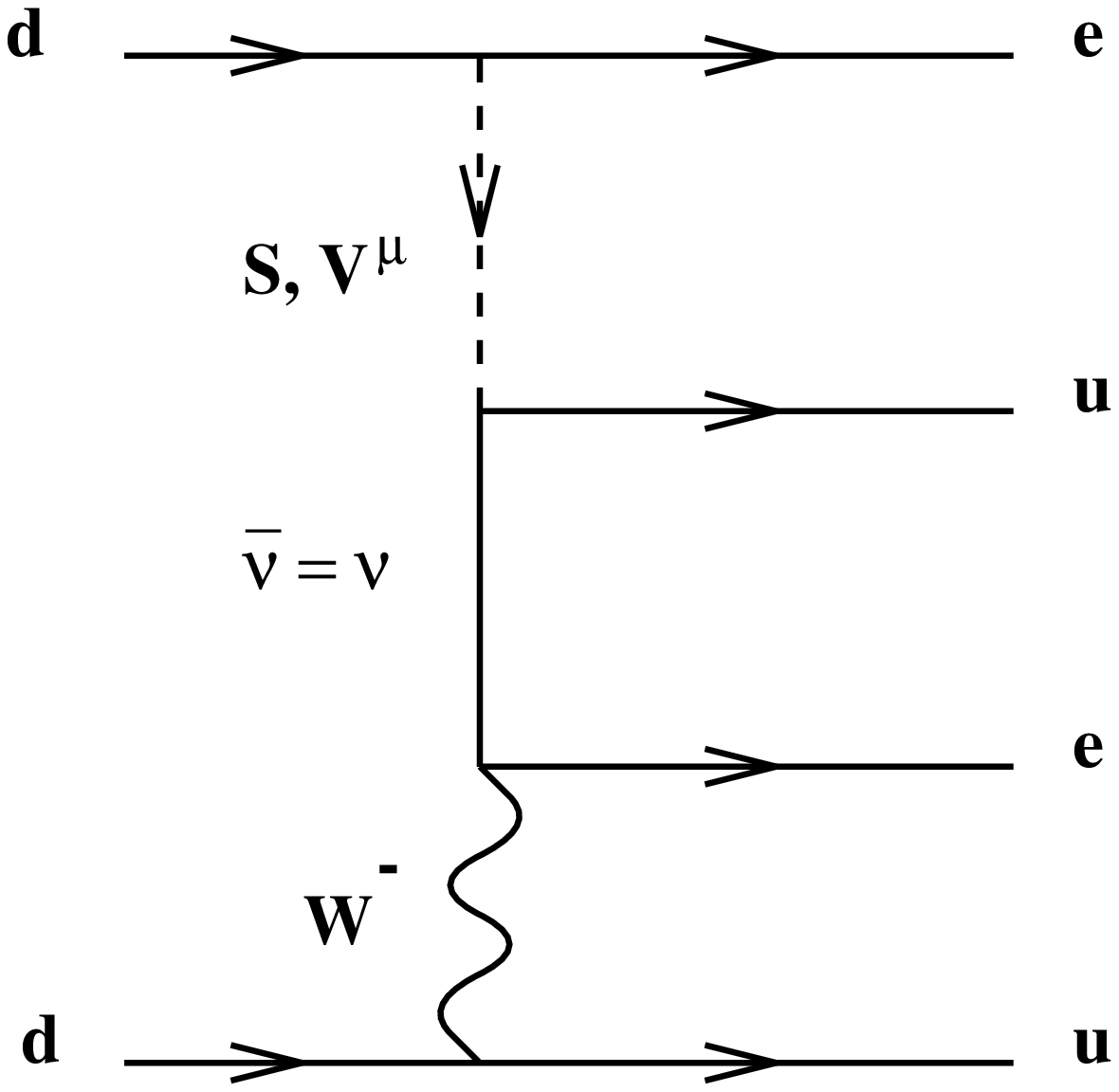}}
\vskip5mm
{\bf Fig. 3} {\it Examples of Feynman graphs for $0\nu\beta\beta$ decay 
within LQ models. $S$ and $V^{\mu}$ stand symbolically for scalar 
and vector LQs, respectively (from [Hir96a]).}
\end{figure}

To consider LQ phenomenology in a model-independent fashion one 
usually follows some general principles in constructing the Lagrangian 
of the LQ interactions with the standard model fields. In order to 
obey the stringent constraints from (c1) helicity-suppressed 
$\pi \rightarrow e\nu$ decay, from (c2) FCNC processes and 
from (c3) proton stability, the following assumptions are commonly adopted: 
(a1) LQ couplings are chiral, (a2) LQ couplings are generation 
diagonal, and (a3) there are no diquark couplings. 

Recently, however, it has been pointed out \cite{hir96a} that possible 
LQ-Higgs interactions spoil assumption (a1): Even if one assumes 
LQs to be chiral at some high energy scale, LQ-Higgs interactions 
introduce after electro-weak symmetry breaking mixing between 
LQ states with different chirality. Since there is no fundamental 
reason to forbid such LQ-Higgs interactions, it seems difficult 
to get rid of the unwanted non-chiral interactions in LQ models. 

In such LQ models there appear contributions to $0\nu\beta\beta$ 
decay via the Feynman graphs of Fig. 3. Here, $S$ and $V^{\mu}$ 
stand 
symbolically for scalar and vector LQs, respectively. 
The half--life for $0\nu\beta\beta$ decay arising from leptoquark 
exchange is given by \cite{hir96a}

\be
T_{1/2}^{0\nu}=|M_{GT}|^2 \frac{2}{G_F^2}[\tilde{C}_1a^2+C_4 b_R^2
+2 C_5 b_L^2].
\ee

with $a=\frac{\epsilon_S}{M_S^2}+\frac{\epsilon_{V}}{M_V^2}$, 
$b_{L,R}=\frac{\alpha_{S}^{(L,R)}}{M_S^2}+\frac{\alpha_V^{(L,R)}}{M_V^2}$,
$\tilde{C}_1=C_1 \Big(\frac{{\cal M}_1^{(\nu)}/(m_e R)}{M_{GT}-
\alpha_2 M_F}
\Big)^2$.

For the definition of the $C_n$ see \cite{74} and for
the calculation 
of the
matrix element ${\cal M}_{1}^{(\nu)}$ see \cite{hir96a}.
This allows to deduce information on leptoquark masses and leptoquark--Higgs
couplings (see section 3.2).

\subsection{Special Relativity and Equivalence Principle}
Special relativity 
and the equivalence principle can be considered as the most 
basic foundations of the theory of gravity. 
Many experiments already have tested these principles to a very high 
level of
accuracy \cite{rel} for ordinary matter - generally for 
quarks and leptons of the first
generation. These precision tests of 
local Lorentz invariance -- violation of the equivalence 
principle should produce a similar effect \cite{will} -- probe for any 
dependence of the (non--gravitational) laws of physics on a laboratory's 
position, orientation or velocity relative to some preferred frame of
reference, such as the frame in which the cosmic microwave background is 
isotropic.  

A typical feature of the violation of local Lorentz invariance (VLI)
is that different species of matter have a characteristical 
maximum attainable speed.
This can be tested in various sectors of the standard model
through vacuum Cerenkov radiation \cite{gasp}, photon decay \cite{cole},
neutrino oscillations \cite{glash,nu1,nu2,hal,nu3} and $K-$physics
\cite{hambye,vepk}. These arguments can be extended
to derive new constraints from neutrinoless double
beta decay \cite{KPS}. 

The equivalence principle implies that spacetime is described by
unique operational geometry and hence universality of the gravitational 
coupling for all species of matter. In the recent years there
have been attempts to constrain a possible amount of 
violation of the equivalence principle (VEP) in the neutrino sector
from neutrino oscillation experiments \cite{nu1,nu2,hal,nu3}.
However, these bounds do not apply when the gravitational and the
weak eigenstates have small mixing. In a recent paper \cite{KPS} 
a generalized formalism of the neutrino sector has been given to test the VEP
and it has been shown that neutrinoless double beta decay also constrains the 
VEP. VEP implies different neutrino species to suffer from  
different gravitational potentials while propagating through the 
nucleus and hence the effect of different eigenvalues doesn't cancel
for the same effective momentum. 
The main result is that neutrinoless double beta decay can constrain
the amount of VEP even when the mixing angle is zero, {\it i.e.},
when only the weak equivalence principle is violated, for which 
there does not exist any bound at present.

\section{Double Beta Decay Experiments: Present Status and Results}

\subsection{Present Experimental Status}
Fig. 4 shows an overview over measured 
$0\nu\beta\beta$ half--life limits and deduced mass limits. The largest 
sensitivity for $0\nu\beta\beta$ decay is obtained at present by active source 
experiments (source=detector), in particular $^{76}$Ge \cite{KK1,KK2}.

\begin{figure}
\parbox{10cm}{
\vspace*{-3cm}
\epsfxsize9cm
\epsffile{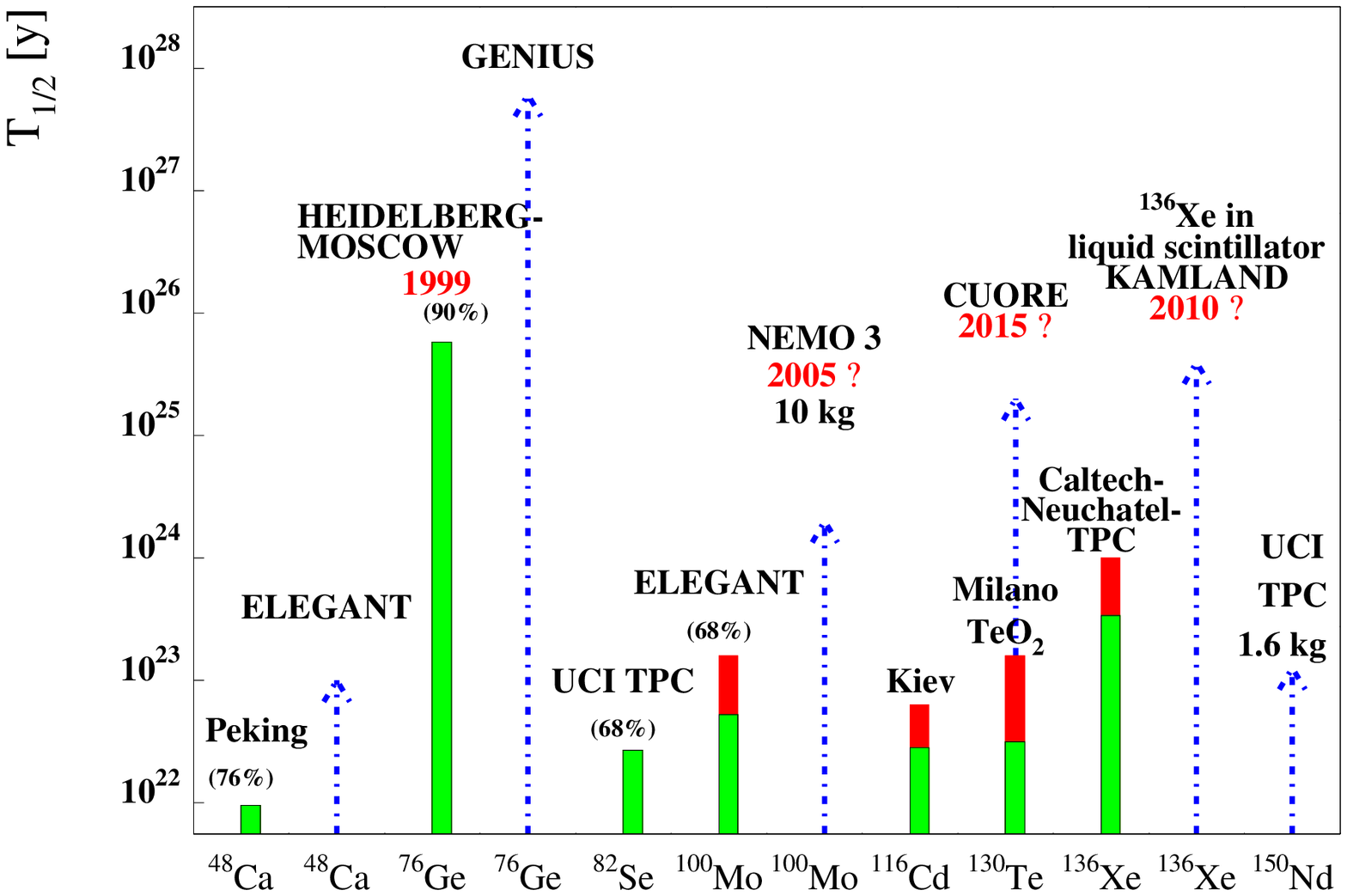}
\vspace*{-5cm}
}
\parbox{10cm}{
\epsfxsize9cm
\epsffile{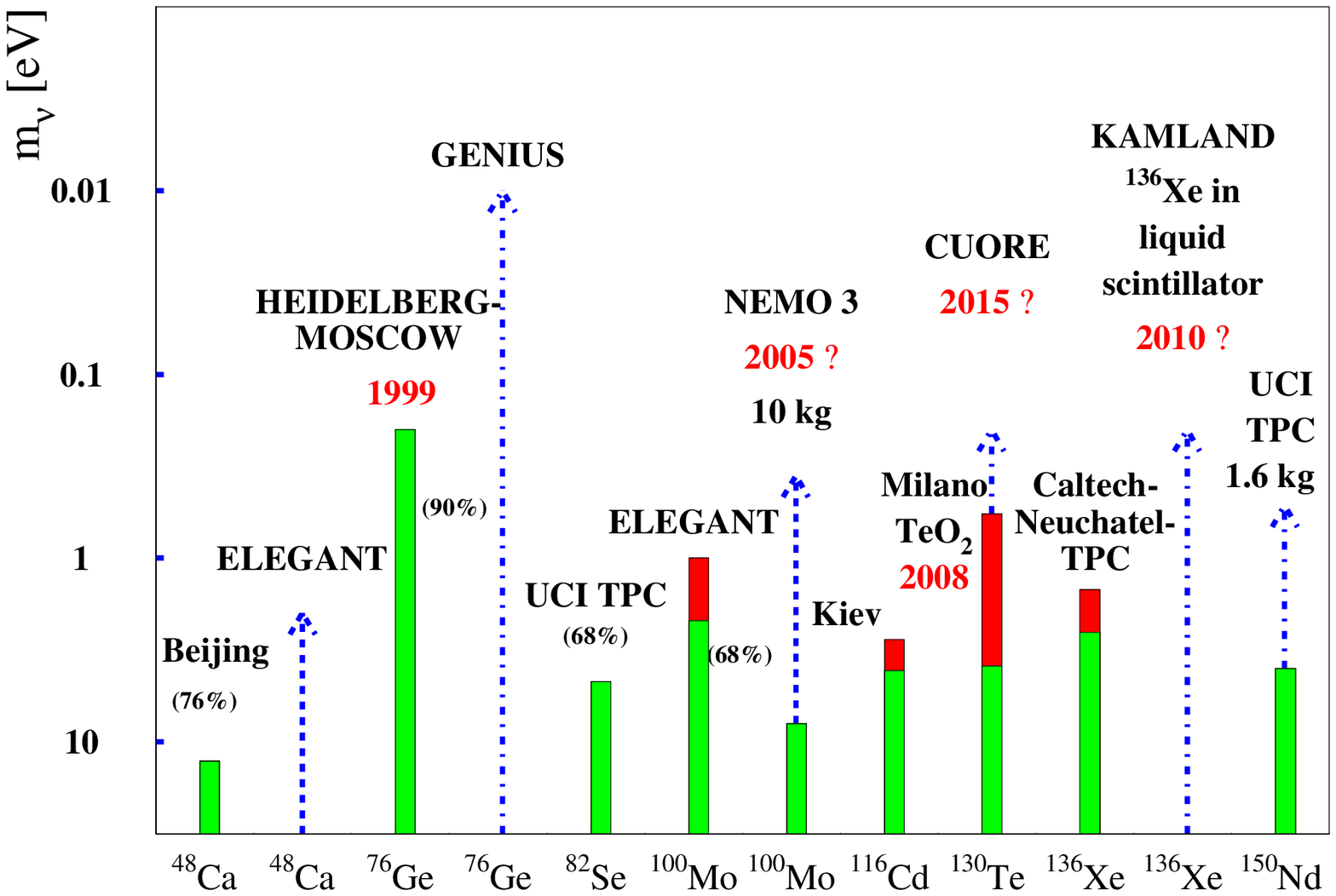}
}
\vspace*{-3cm}

{\bf Fig. 4} {\it 
Present situation, 1999, and expectation for the near future
and beyond, of 
the most promising $\beta\beta$-experiments concerning accessible half life 
(a) and neutrino mass limits (b). The light-shaded parts of the bars 
correspond 
to the present status, the dark parts of the bars to 
expectations for running experiments, dashed lines to 
experiments under construction and dash-dotted lines to proposed experiments.}
\end{figure}

Only a few of the present most sensitive experiments may probe the 
neutrino mass 
in the next years into
the sub--eV region, the 
Heidelberg--Moscow experiment being the by far 
most advanced and most sensitive one, see Fig. 4b. 
No one of them will pass, however,  below $\sim 0.1$ eV (see section
4.1).
A detailed discussion of the various experimental
possibilities can be found in \cite{1,tren,Kla95b,Kla99a}. 
A useful listing of existing 
data from the 
various $\beta\beta$ emitters is given in \cite{83}.

\subsection{Present limits on beyond standard model parameters}      
The sharpest limits from $0\nu\beta\beta$ decay are presently coming from
the Heidelberg--Moscow experiment \cite{84,KK2,Kla99a,Bau99a}. 
They will be given in the following.
With five 
enriched (86\% of $^{76}$Ge) detectors of a total mass of 11.5 kg 
taking data in the Gran Sasso underground laboratory, and with a background
of at present 0.06 counts/kg year keV, 
the experiment has reached its final 
setup and is now 
exploring the sub--eV range for the mass of the electron neutrino.
Fig. 5 shows the spectrum taken in a measuring time of 24  kgy with pulse 
shape analysis..

\subsection*{Half--life of neutrinoless double beta decay}
The deduced half--life limit for $0\nu\beta\beta$ decay is using the method 
proposed by \cite{PDG98} 

\be
T^{0\nu}_{1/2} > 5.7 \cdot 10^{25} y \hspace{2mm}(90\% C.L.)
\ee
\be
\hskip8mm     > 2.5 \cdot 10^{26} y \hspace{2mm}(68 \% C.L.).
\ee

\subsection*{Neutrino mass}

\begin{figure}[!t]
\hspace*{15mm}
\epsfxsize=90mm
\epsfbox{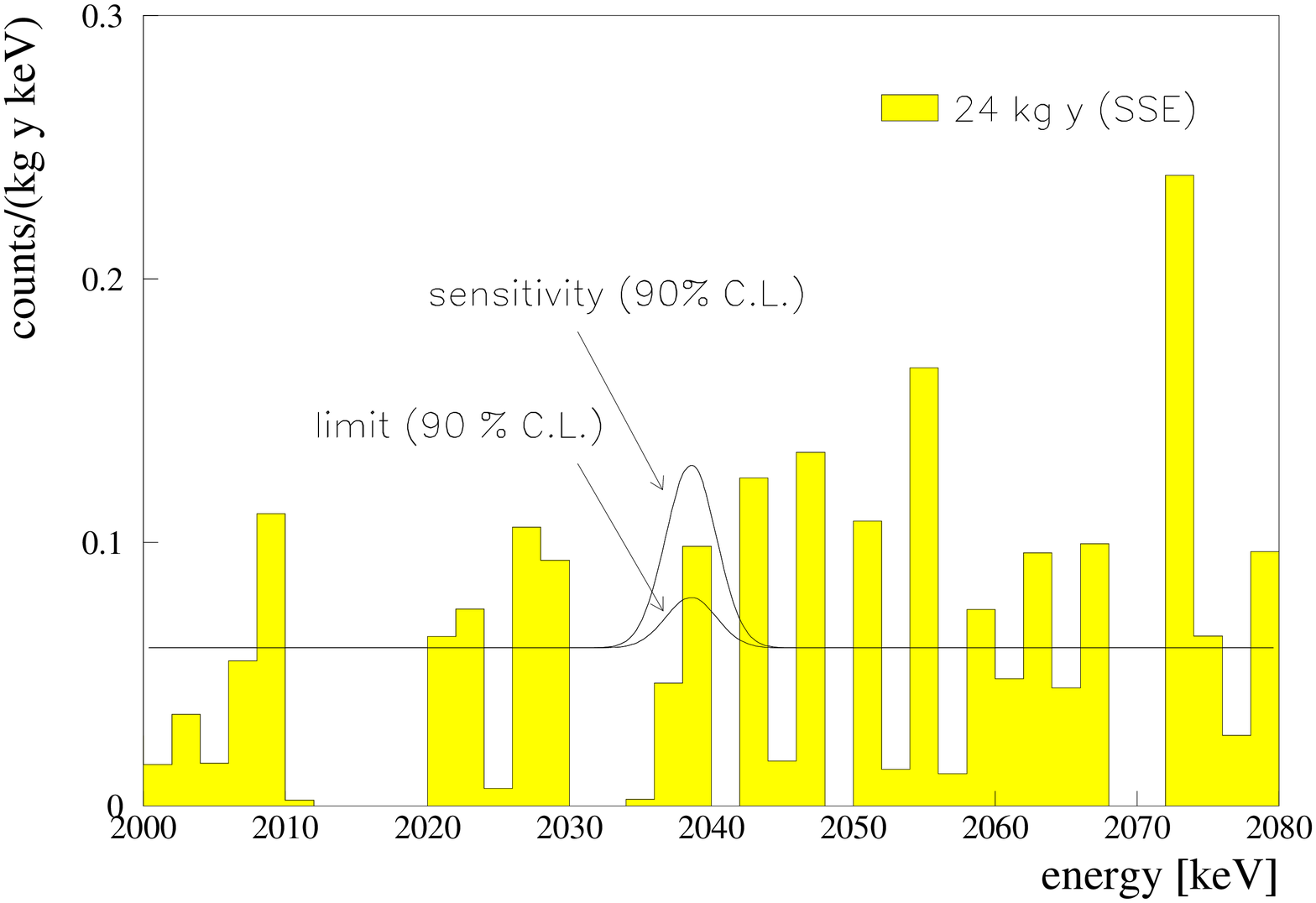}

\noindent
{\bf Fig. 5} {\it Integral spectrum in the region of interest after 
subtraction of the first 200 days of measurement of each detector, 
leaving 24 kg y of measuring time with pulse shape analysis.
The two solid
 curves correspond to the signal excluded
with $90 \% C.L.$ and to the sensitivity defined by \cite{Fel98} of the experiment ($90 \% C.L.$) .
They correspond to $T^{0\nu}_{1/2} > 5.7 \cdot 10^{25}$ y
and  $T^{0\nu}_{1/2} > 1.6 \cdot 10^{25}$ y, respectively.
(from \cite{Bau99a})
}
\end{figure}

{\it Light neutrinos:} The deduced upper limit of an (effective) electron
neutrino 

Majorana mass is, with the matrix element from \cite{29}
\be
\langle m_{\nu} \rangle < 0.20 eV \hspace{2mm}(90\% C.L.)
\ee
\be
\hskip10mm < 0.10 eV \hspace{2mm}(68 \% C.L.)
\ee

This is the sharpest limit for a Majorana mass of the electron neutrino so
far. With these values the Heidelberg--Moscow experiment starts to take 
striking influence on presently discussed neutrino mass scenarios, which arose 
in connection with the recent Superkamiokande results on solar and atmospheric
neutrinos. We mention a few examples:

The new $0\nu\beta\beta$ result excludes already now simultaneous 3$\nu$
solutions for hot dark matter, the atmospheric neutrino problem and the small
mixing angle MSW solution \cite{Adh98}. This means that Majorana neutrinos 
are ruled out, if the small mixing angle solution of the solar neutrino
problem is borne out -- if we insist on neutrinos as hot dark matter
candidates. According to \cite{Min97} degenerate neutrino mass schemes for hot 
dark matter, solar and atmospheric anomalies and CHOOZ anre already now 
excluded (with 68 \% C.L.) for the small {\it and} 
large mixing angle MSW solutions
(without unnatural finetuning). If starting from recent dark matter models 
\cite{Pri98} including in addition to cold and hot dark matter also a 
cosmological constant $\Lambda \neq 0$, these conclusions remain also valid, 
except for the large angle solution which would not yet be excluded by
$0\nu\beta\beta$ decay (see \cite{Kla99b}).

According to \cite{Bar98} simultaneous 3$\nu$ solutions of solar and 
atmospheric neutrinos, and LSND and CHOOZ (no hot dark matter!) predict
$\langle m_{\nu} \rangle\simeq 1.5 eV$ for the degenerate case 
($m_i \simeq 1 eV$) and
$\langle m \rangle \simeq 0.14 eV$ for the hierarchical case.    
This means {\it both} cases are practically excluded already by the present 
Heidelberg--Moscow result. A model producing the neutrino masses based on a
heavy scalar triplet instead of the seesaw mechanism derives from the solar
small angle MSW allowed range of mixing, and accomodating the atmospheric 
neutrino problem, $\langle m_{\nu} \rangle$ =0.17-0.31 eV \cite{Ma99}. Also
this model is already excluded with 68 \% C.L., including an uncertainty 
of a factor of 2 in the nuclear matrix elements. Looking into 4-neutrino 
scenarios, according to \cite{Giu99} there are only two schemes with
four neutrino mixing that can accomodate the results of {\it all} 
neutrino oscillation experiments (including LSND). In the first of the schemes,
where $m_1 < m_2 \ll m_3 < m_4$, with 
solar (atmospheric) neutrinos oscillating between $m_3$ and $m_4$ ($m_1$ and 
$m_2$), and 
$\Delta m^2_{LSND}= \Delta m_{41}^2$,
the HEIDELBERG--MOSCOW $0\nu\beta\beta$ bound excludes \cite{Giu99} the
small mixing angle MSW solution of the solar neutrino problem, for both 
$\nu_e \rightarrow \nu_{\tau}$, and $\nu_e \rightarrow \nu_s$ transitions. 
Including recent astrophysical data yielding $N_{\nu}^{BBN}\leq 3.2$ 
(95 \% C.L.) \cite{Bur99}, the oscillations of solar neutrinos occur mainly
in the $\nu_e \rightarrow \nu_s$ channel, and {\it only} the small angle 
solutions is allowed by the fit of the solar neutrino data \cite{Bah98,Fuk99}.
This means that $0\nu\beta\beta$ excludes the whole first scheme.

In the second scheme $m_1 < m_2 \ll m_3 < m_4$, with solar (atmospheric) 
neutrinos oscillating between $m_1$ and $m_2$ ($m_3$ and $m_4$), 
the present neutrino 
oscillation experiments indicate an effective Majorana mass of 
$7 \cdot 10^{-4} eV \leq |\langle m \rangle| \leq 2 \cdot 10^{-2} eV$. This
could eventually be measured by GENIUS (see below). For a similar recent 
analysis see \cite{Bil99}. For further detailed analyses of neutrino mass 
textures in the light of present and future
neutrino experiments including double beta 
decay we refer to \cite{Kla99b}. 

{\it Superheavy neutrinos:}     
For a superheavy {\it left}--handed neutrino we deduce 
\cite{79,14,Bel98} exploiting the 
mass dependence of the matrix 
element (for the latter 
see \cite{28}) a lower limit (see also Fig. 11)
\be
\langle m_{H} \rangle \ge 100 TeV.
\ee











\subsection*{Right--handed W boson}
For the right--handed W boson we obtain a lower limit of 
\be
m_{W_R} \ge 1.6 TeV
\ee
(see \cite{11,KKP}).

\subsection*{SUSY parameters -- R--parity breaking and sneutrino mass}
The constraints on the parameters of the minimal supersymmetric standard model
 with explicit R--parity violation deduced \cite{6,hir96c,hir96} 
from the $0\nu\beta\beta$
half--life limit are more stringent than those from other
low--energy processes and from the largest high energy
accelerators. The limits are 
\be
\lambda^{'}_{111} \leq 3.9 \cdot 10^{-4} \Big(\frac {m_{\tilde{q}}}{100 GeV} 
\Big)^2 \Big(\frac {m_{\tilde{g}}}{100 GeV} \Big)^{\frac{1}{2}}
\ee
with  $m_{\tilde{q}}$ and  $m_{\tilde{g}}$ denoting squark and gluino masses,
respectively, and with the assumption $m_{\tilde{d_R}} \simeq m_{\tilde{u}_L}$.
This result is important for the discussion of new physics in the connection
with the high--$Q^2$ events seen at HERA. It excludes the possibility of 
squarks of first generation (of R--parity violating SUSY) being produced in the
high--$Q^2$ events \cite{Cho97,Alt97,Hir97b}.

We find further \cite{hir96} 
\be
\lambda_{113}^{'}\lambda_{131}^{'}\leq 1.1 \cdot 10^{-7}
\ee
\be
\lambda_{112}^{'}\lambda_{121}^{'}\leq 3.2 \cdot 10^{-6}.
\ee 
For the $(B-L)$ violating sneutrino mass $\tilde{m}_{M}$ the following limits 
are obtained \cite{Hir97a}
\ba{rconv2}
\tilde{m}_M &\leq& 2 \Big(\frac{m_{SUSY}}{100 GeV}\Big)^{\frac{3}{2}}GeV,
\hskip5mm \chi \simeq \tilde{B}\\
\tilde{m}_M &\leq& 11 \Big(\frac{m_{SUSY}}{100 GeV}\Big)^{\frac{7}{2}}GeV,
\hskip5mm \chi \simeq \tilde{H}
\ea
for the limiting cases that the lightest neutralino is a pure Bino $\tilde{B}$,
as suggested by the SUSY solution of the dark matter problem \cite{Jun96},
or a pure Higgsino. Actual values for $\tilde{m}_M$ for other choices of the
neutralino composition should lie in between these two values.
 
Another way to deduce a limit on the `Majorana' sneutrino mass $\tilde{m}_M$
is to start from the experimental neutrino mass limit, since the sneutrino 
contributes to the Majorana neutrino mass $m_M^{\nu}$ at the 1--loop level
proportional to $\tilde{m}^2_M$. This yields under some assumptions
\cite{Hir97a}
\be
\tilde{m}_{M_{(i)}} \leq (60-125) \Big(\frac{m^{exp}_{\nu(i)}}{1 eV}\Big) ^{1/2}
MeV
\ee

Starting from the mass limit determined for the electron neutrino  by 
$0\nu\beta\beta$ decay this leads to 
\be
\tilde{m}_{M_{(e)}} \leq 22 MeV    
\ee
This result is somewhat dependent on neutralino masses and mixings. 
A non--vanishing `Majorana' sneutrino mass would result in new processes 
at future colliders, like sneutrino--antisneutrino oscillations.
Reactions at the Next Linear Collider (NLC) like the SUSY analog to inverse
neutrinoless double beta decay $e^-e^-\rightarrow \chi^-\chi^-$ (where $\chi^-$
denote charginos) or single sneutrino production, e.g. by 
$e^-\gamma \rightarrow \tilde{\nu}_e \chi^-$ could give information on the 
Majorana sneutrino mass, also. This is discussed by \cite{Hir97,Hir97a,Kolb1}.
A conclusion is that future
accelerators can give information on second and third generation sneutrino
Majorana masses, but for first generation sneutrinos cannot compete with
$0\nu\beta\beta$--decay.


\subsection*{Compositeness}
Evaluation of the $0\nu\beta\beta$ half--life limit assuming 
exchange of excited
Majorana neutrinos $\nu^*$ yields for the mass of the
 excited neutrino a lower bound of \cite{Pan97,Tak97}. 
\be
m_{N} \geq 3.4 m_W
\ee
for a coupling of order \cal{O}(1) and $\Lambda_c \simeq m_N$. Here,
$m_W$ is the W--boson mass.

\subsection*{Leptoquarks}
Assuming that either scalar or vector leptoquarks contribute
to $0\nu\beta\beta$ decay, the following constraints on the 
effective LQ parameters  (see section 2.7) can be derived \cite{hir96a}:
\ba{dbd_constraint}
\epsilon_I \leq 2.8 \times 10^{-9}
\left(\frac{M_I}{100\mbox{GeV}}\right)^2, \\
\alpha_I^{(L)} \leq 3.5 \times 10^{-10}
\left(\frac{M_I}{100\mbox{GeV}}\right)^2, \\
\alpha_I^{(R)} \leq 7.9  \times 10^{-8}
\left(\frac{M_I}{100\mbox{GeV}}\right)^2.
\ea

%

Since the LQ mass matrices appearing in $0\nu\beta\beta$ 
decay are ($4\times4$) 
matrices \cite{hir96a}, it is difficult to solve their diagonalization 
in full generality algebraically. However, if one assumes that only 
one LQ-Higgs coupling is present at a time, the (mathematical) problem is 
simplified greatly and one can deduce from, for example, 
eq. 40 that either 
the LQ-Higgs coupling must be smaller than $\sim 10^{-(4-5)}$ or there can not 
be any LQ with e.g. couplings of electromagnetic strength with masses below
$\sim 250 GeV$. These bounds from $\beta\beta$ decay are of interest in 
connection with recently discussed evidence for new physics from HERA
\cite{Hew97,Bab97,Kal97,Cho97}. Assuming that actually leptoquarks have
been produced at HERA, double beta decay (the Heidelberg--Moscow experiment)
would allow to fix the leptoquark--Higgs coupling to a few $10^{-6}$
\cite{Hir97b}. It may be noted, that after the first 
consideration of leptoquark--Higgs coupling in \cite{hir96a} recently
Babu et al. \cite{Bab97b} noted that taking into account 
leptoquark--Higgs coupling reduces the leptoquark mass lower bound deduced
by TEVATRON -- making it more consistent with the value of 200 GeV 
required by 
HERA.

\subsection*{Special Relativity and Equivalence Principle}
{\it Violation of Lorentz invariance (VLI):} The bound obtained from the 
Heidelberg--Moscow experiment is
\be
\delta v < 4 \times 10^{-16}~~~~ {\rm for}~~~ \theta_v=\theta_m =0
\ee
where $\delta v=v_1-v_2$ is the measure of VLI in the neutrino sector.
$\theta_v$ and $\theta_m$ denote the velocity mixing angle and the weak 
mixing angle, respectively.
In Fig. 6 (from \cite{KPS}) the bound implied by double beta decay is 
presented for the entire
range of $sin^2(2 \theta_v)$, and compared with bounds obtained from
neutrino oscillation experiments (see \cite{hal}).

\begin{figure}[!t]
\epsfysize=80mm
\hspace*{5mm}
\epsfbox{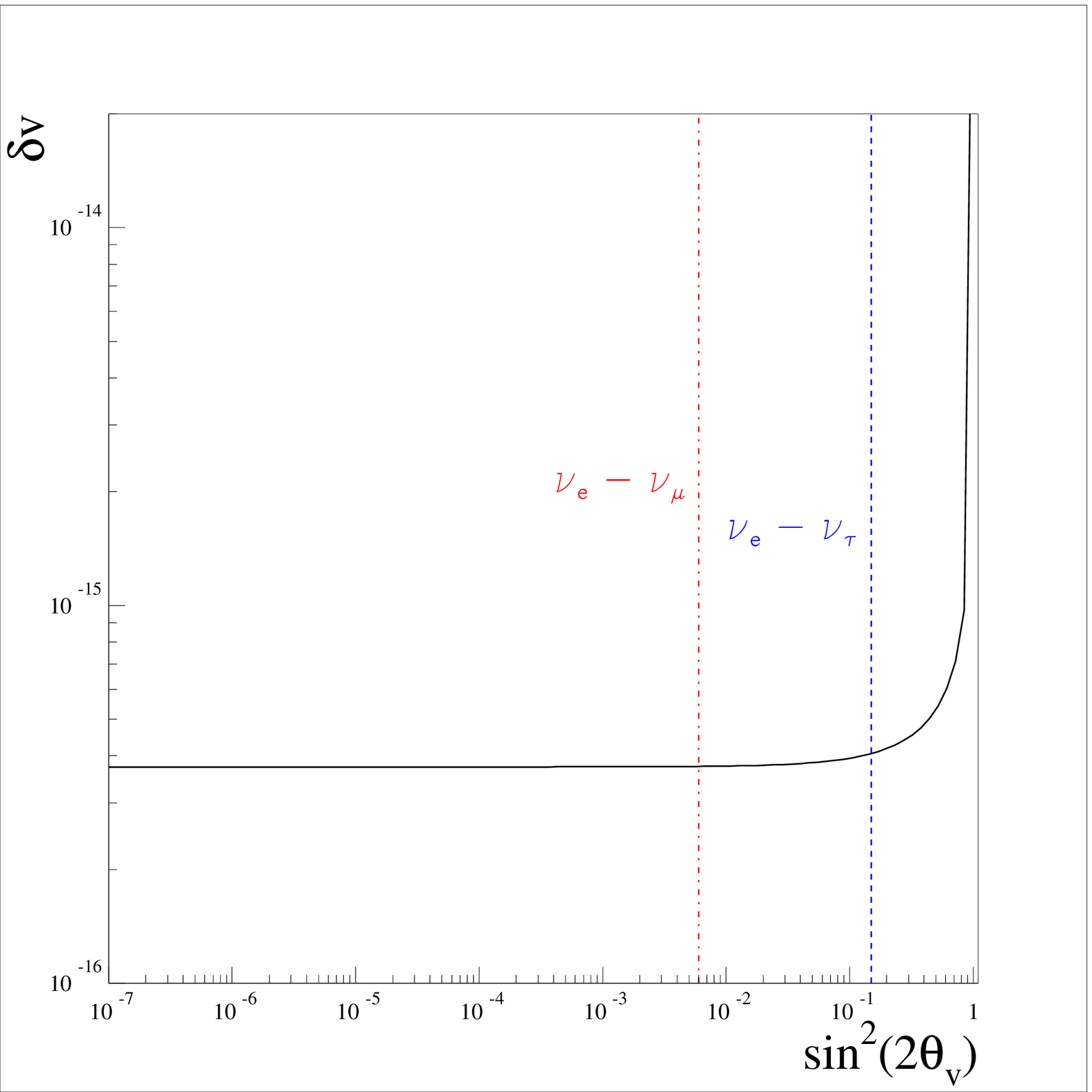}
\vspace*{5mm}\\
\noindent
\bf{Fig. 6}
{\it Double beta decay bound (solid line)
on violation of Lorentz invariance 
in the neutrino sector, excluding the region to the upper left. 
Shown is a double logarithmic plot 
in the $\delta v$--$\sin^2(2 \theta)$ parameter space. 
The bound becomes most stringent for the
small mixing region, which has not been constrained from any
other experiments. For comparison the bounds obtained from neutrino oscillation
experiments (from \protect{\cite{hal}})
in the $\nu_{e} - \nu_{\tau}$ (dashed lines) and in the
$\nu_e - \nu_\mu$ (dashed-dotted lines) channel, excluding the region to the 
right, are shown (from \protect{\cite{KPS}}).}
\end{figure}
\nopagebreak
{\it Violation of equivalence principle (VEP):}
Assuming only violation of the weak equivalence principle, there does not 
exist any bound on the amount of VEP. It is this region of the parameter space
which is most restrictively bounded by neutrinoless double beta decay.
In a linearized theory the gravitational part of the Lagrangian to first order
in a weak gravitational field $g_{\mu\nu}=\eta_{\mu\nu}+    h_{\mu\nu}$
($h_{\mu\nu}= 2\frac{\phi}{c^2}  {\mbox diag}(1,1,1,1)$)
can be written as ${\cal L} = -\frac{1}{2}(1+g_i)h_{\mu\nu}T^{\mu\nu}$,
where  $T^{\mu\nu}$  is the  stress-energy  in the  gravitational
eigenbasis. In the presence of VEP the $g_i$ may differ.
We obtain \cite{KPS} the following bound from the Heidelberg--Moscow 
experiment, for $\theta_v=\theta_m=0$:
  \ba{99}
\phi \delta g &<& 4 \times 10^{-16} ~ ({\rm for~} \bar{m}<13 
{\rm eV})\nn \\
\phi \delta g &<& 2 \times 10^{-18} ~ ({\rm for~} \bar{m}<0.08 
{\rm eV}).
\ea
Here $\bar{g}=\frac{g_1+g_2}{2}$ can be considered as the standard 
gravitational coupling, for which the equivalence principle applies.
$\delta g=g_1 - g_2$.
The bound on the VEP thus, unlike the one for VLI, will depend on the choice
for the Newtonian potential $\phi$.

\subsection*{Half--life of $2\nu\beta\beta$ decay}
The Heidelberg--Moscow experiment 
produced for the first time a high statistics $2\nu\beta\beta$
spectrum ($\gg$ 20000 counts, to be compared with the 40 counts on which the 
first detector observation of $2\nu\beta\beta$ decay by \cite{Ell87} 
(for the decay of $^{82}$Se) had to rely). 
The deduced half--life is \cite{HM97}
\be
T^{2\nu}_{1/2} = (1.77^{+0.01}_{-0.01}(stat.)^{+0.13}_{-0.11}(syst.)) 
\cdot 10^{21} y
\ee
This result brings $\beta\beta$ research for the first time into the region 
of `normal' nuclear spectroscopy and allows for the first time statistically
reliable investigation of Majoron--accompanied decay modes.

\subsection*{Majoron--accompanied decay}
From simultaneous fits of
the $2\nu$ spectrum and one selected Majoron mode, experimental limits 
for the half--lives of the decay modes of
the newly introduced Majoron models \cite{72} are given
for the first time \cite{71,HM96}.

The small matrix elements and phase spaces for these modes 
\cite{71,75} already determined that these 
modes by far cannot be seen
in experiments of the present sensivity if we assume typical values for the
neutrino--Majoron coupling constants around $\langle g \rangle = 10^{-4}$. 

\section{Double Beta Experiments: Future Perspectives -- 
the GENIUS Project}
\subsection{The known experiments and proposals}
Figs. 4a,b show in addition to the present status 
the future perspectives of the main existing 
$\beta\beta$ decay experiments and includes some ideas for the future
which have been published.
The best presently existing limits besides the HEIDELBERG-MOSCOW 
experiment (filled bars in Fig. 4),
have been obtained with the isotopes: 
$^{48}$Ca \cite{87}, 
$^{82}$Se \cite{88}, 
$^{100}$Mo \cite{89}, 
$^{116}$Cd \cite{90},
$^{130}$Te \cite{91},
$^{136}$Xe \cite{92} and
$^{150}$Nd \cite{93}.
These and other double beta decay setups presently under construction or 
partly in operation 
such as NEMO \cite{94,Bar97}, 
the Gotthard $^{136}$Xe TPC experiment \cite{95}, 
the $^{130}$Te cryogenic experiment \cite{91},
a new ELEGANT $^{48}$Ca experiment using 30 g of $^{48}$Ca \cite{96},
a hypothetical experiment with an improved UCI TPC \cite{93} assumed to use 1.6 kg of $^{136}$Xe, 
 etc., will not reach or exceed the $^{76}$Ge limits.
The goal 0.3 eV aimed at for the year 2004 by the NEMO experiment 
(see \cite{98,Bar97}
and Fig. 4) 
may even be very optimistic if claims about the effect of proton-neutron 
pairing on the $0\nu\beta\beta$ nuclear matrix elements by 
\cite{Pan96} will
turn out to be true, and also if the energy resolution will not be improved
considerably 
(see Fig. 1 in \cite{83}). 
Therefore, the conclusion given by \cite{Bed97c} concerning the
future SUSY potential of NEMO has no serious basis. 
As pointed out by Raghavan \cite{97}, even use of an 
amount of about 200 kg of 
enriched $^{136}$Xe or 2 tons of natural Xe added to the scintillator of the 
KAMIOKANDE detector 
or similar amounts added to BOREXINO (both primarily devoted to solar neutrino 
investigation) 
would hardly lead to a sensitivity larger 
than the present $^{76}$Ge experiment.
This idea is going to be realized at present by the KAMLAND
experiment \cite{Suz97}. 

It is obvious from Fig. 4 that {\it none}
of the present experimental approaches, or plans or even vague ideas has a
chance to surpass the border of 0.1 eV for the neutrino mass to lower values
(see also \cite{Nor97}).
At present there is only one way visible to reach the domain of lower 
neutrino masses,
suggested by \cite{KK1} and meanwhile investigated 
in some
detail concerning its experimental realization and physics potential in
\cite{Kla97d,Hel97,KK2,KK3}.

\subsection{Genius -- A Future Large Scale Double Beta and Dark Matter
Experiment}

The idea of GENIUS is to use a large amount of `naked' enriched 
{\bf GE}rmanium detectors in liquid {\bf NI}trogen as shielding in an 
{\bf U}nderground {\bf S}etup. Use of 1 (in an extended version 10) tons of
enriched $^{76}$Ge will increase the source strength largely, removing all
material from the vicinity of the detectors and shielding by liquid nitrogen
will lead to a drastic background reduction compared to the present level.
That Ge detectors can be operated in liquid nitrogen has been demonstrated
recently in the Heidelberg low level laboratory \cite{Hel97,Bau98}. 

\begin{figure}[!t]
\epsfxsize10cm
\epsffile{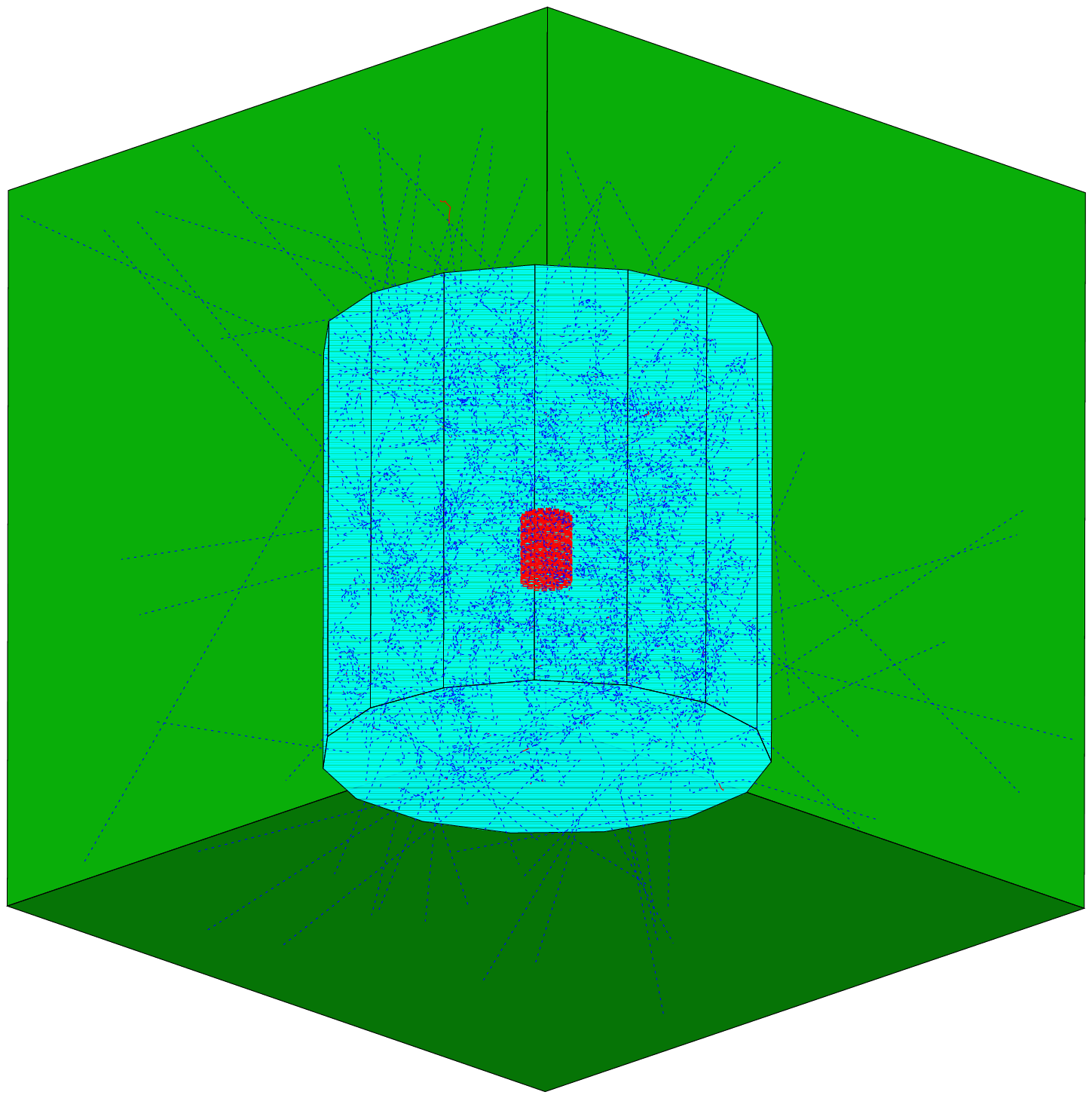}
{\bf Fig. 7}
{\it Simplified model of the GENIUS experiment: 288 enriched
$^{76}$Ge detectors
with a total of one ton mass in the center of a 12 m high liquid nitrogen 
tank with 12 m diameter; GEANT Monte Carlo simulation of 1000  2.6 MeV 
photons randomly
distributed in the nitrogen is also shown.}
\end{figure}

\subsubsection{Realization and Sensitivity of GENIUS}
A simplified model of GENIUS is shown in Fig. 7 consisting of about 300
enriched $^{76}Ge$ detectors with a total of one ton mass in the center of a 
12 m high liquid nitrogen tank with 12 m diameter. The results
of Monte Carlo simulations, using the CERN GEANT code, of the background
\cite{Hel97,Bau98}, 
starting from purity levels of the nitrogen being in general 
an order of magnitude less stringent than those already achieved in the CTF for the
BOREXINO experiment, yield for the count rate 
in the region of interest for neutrinoless double beta decay is 0.04 
counts/(keV y t). Below 100 keV the background 
count rate is about 10 counts/(keV y t).
Two neutrino double beta decay would dominate the spectrum with $4\cdot 10^6$
events per year (for details see \cite{KK2,KK3,Bau98}).



Starting from these numbers, a lower half--life limit of
\be
T_{1/2}^{0\nu} \geq 5.8 \cdot 10^{27} \hskip8mm (68 \% C.L.)
\ee
can be reached within one year of measurement (following the 
highly conservative procedure for
analysis recommended by \cite{46}, which has been used also in the 
derivation of the results given in section 3.2, but is not used in the analysis 
of several other $\beta\beta$ experiments). This corresponds 
-- with the matrix
elements of \cite{29} -- to an upper limit on the neutrino mass of

\be
\langle m_{\nu}\rangle \leq 0.02 eV \hskip8mm (68 \% C.L.)
\ee 

The final sensitivity of the experiment can be defined by the limit, 
which would
be obtained after 10 years of measurement assuming zero background. For the one ton experiment this 
would be:
\medskip
\be
T^{0\nu}_{1/2}\quad\ge\quad 6.4 \cdot 10^{28}\, y \quad \mbox{(with
68\% C.L.)}\
\ee
\medskip
and
\medskip
\be
\langle m_{\nu}\rangle \quad \le 0.006 eV \quad \mbox{(with 68\% C.L.)}\
\ee
\medskip

The ultimate experiment could test the $0\nu\beta\beta$ half life of
$^{76}$Ge up to a limit of 5.7$\cdot$10$^{29}$y 
and the neutrino mass down to 2$\cdot$10$^{-3}$eV using
10 tons of enriched Germanium and a measuring time of 10 years.

\subsubsection{The Physics Potential of GENIUS} \hspace*{50mm}\\
{\it Neutrino mass textures and neutrino oscillations:}
GENIUS will allow a large step in sensitivity for probing the neutrino mass. 
It will allow to probe the neutrino mass down to 10$^{-(2-3)}$ eV, and thus 
surpass  the existing neutrino mass experiments by a factor of 50-500.
GENIUS will test the structure of the neutrino mass matrix and thereby also 
neutrino oscillation parameters 
\footnote{The double beta observable, the effective neutrino mass 
(eq. 10), can be expressed 
in terms of the usual neutrino oscillation parameters, once an assumption
on the ratio of $m_1/m_2$ is made. E.g., in the simplest two--generation case
\be
\langle m_{\nu} \rangle=|c_{12}^2 m_1 + s_{12}^2 m_2 e^{2 i \beta}|,
\ee
assuming CP conservation, i.e. $e^{2 i \beta}=\eta=\pm 1$, and 
$c_{12}^2 m_1 << \eta s_{12}^2 m_2$,
\be
\Delta^2_{m_{12}}\simeq m_2^2=\frac{4 \langle m_{\nu} \rangle^2}{1-\sqrt{1-
sin^2 2 \theta}}
\ee
A little bit more general, keeping corrections of the order $(m_1/m_2)$ 
one obtains
\be
m_2=\frac{ \langle m_{\nu} \rangle}{|(\frac{m_1}{m_2})+\frac{1}{2}
(1-\sqrt{1-sin^2 2 \theta})(\pm 1 - (\frac{m_1}{m_2}))|}.
\ee
For the general case see \cite{Kla97d}.}
superior in sensitivity to the best
proposed dedicated terrestrial neutrino oscillation experiments. Even in the 
first stage GENIUS will confirm or rule out degenerate or inverted neutrino 
mass scenarios, discussed in the literature as possible solutions of current 
hints to finite neutrino masses (see \cite{Kla99b,Giu99,Cza99,Vis99}).
If the $10^{-3}$ eV
level is reached, GENIUS will allow to test the large angle and for degenerate
models even the small angle MSW 
solution of the solar neutrino problem. It will also allow to test the 
hypothesis of a shadow world underlying introduction of a sterile neutrino
mentioned in section 2.1.
The figures 8-10 show some examples of this potential (for more
details see \cite{Kla97d,KK1,KK2,KK3,Kla99a}. Fig. 8 compares the
potential of GENIUS with the sensitivity of CHORUS/NOMAD and with the
proposed future experiments NAUSIKAA--CERN and NAUSIKAA--FNAL, looking for 
$\nu_e \leftrightarrow \nu_{\tau}$ oscillations, for different assumptions on
$m_1/m_2$. 
\begin{figure}[!t]
\setlength{\unitlength}{1in}                                                 
\begin{picture}(5,2)
\put(0.0,0.5){\includegraphics{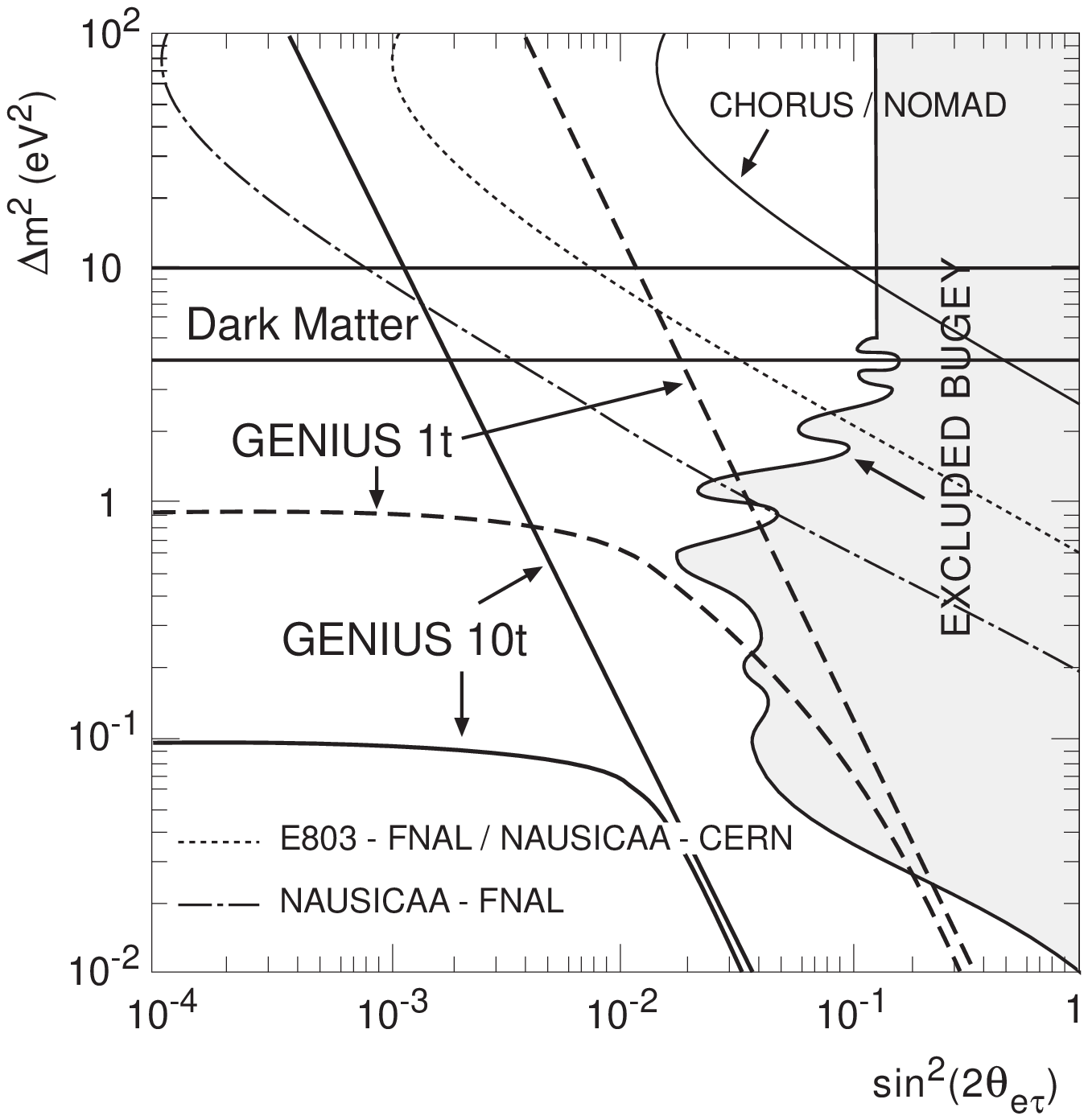}}
\end{picture}                                                                 

\vspace*{40mm}
{\bf Fig. 8}
{\it Current limits and future experimental sensitivity 
on $\nu_e - \nu_{\tau}$ oscillations. The shaded area is currently 
excluded from reactor experiments. The thin line is the estimated 
sensitivity of the CHORUS/NOMAD experiments. The dotted and dash-dotted 
thin lines are sensitivity limits of proposed accelerator experiments, 
NAUSICAA and E803-FNAL [Gon95]. 
The thick lines show the sensitivity of GENIUS (broken line: 
1 t, full line: 10 t), for two examples of mass ratios. The straight lines are 
for the strongly hierarchical case (R=0), while the lines bending to the left 
assume R=0.01.  
(from [Kla97c])}
\end{figure}





\begin{figure}[!t]
\vskip0mm 
\hskip5mm
\epsfxsize=58mm
\epsfysize=58mm
\epsfbox{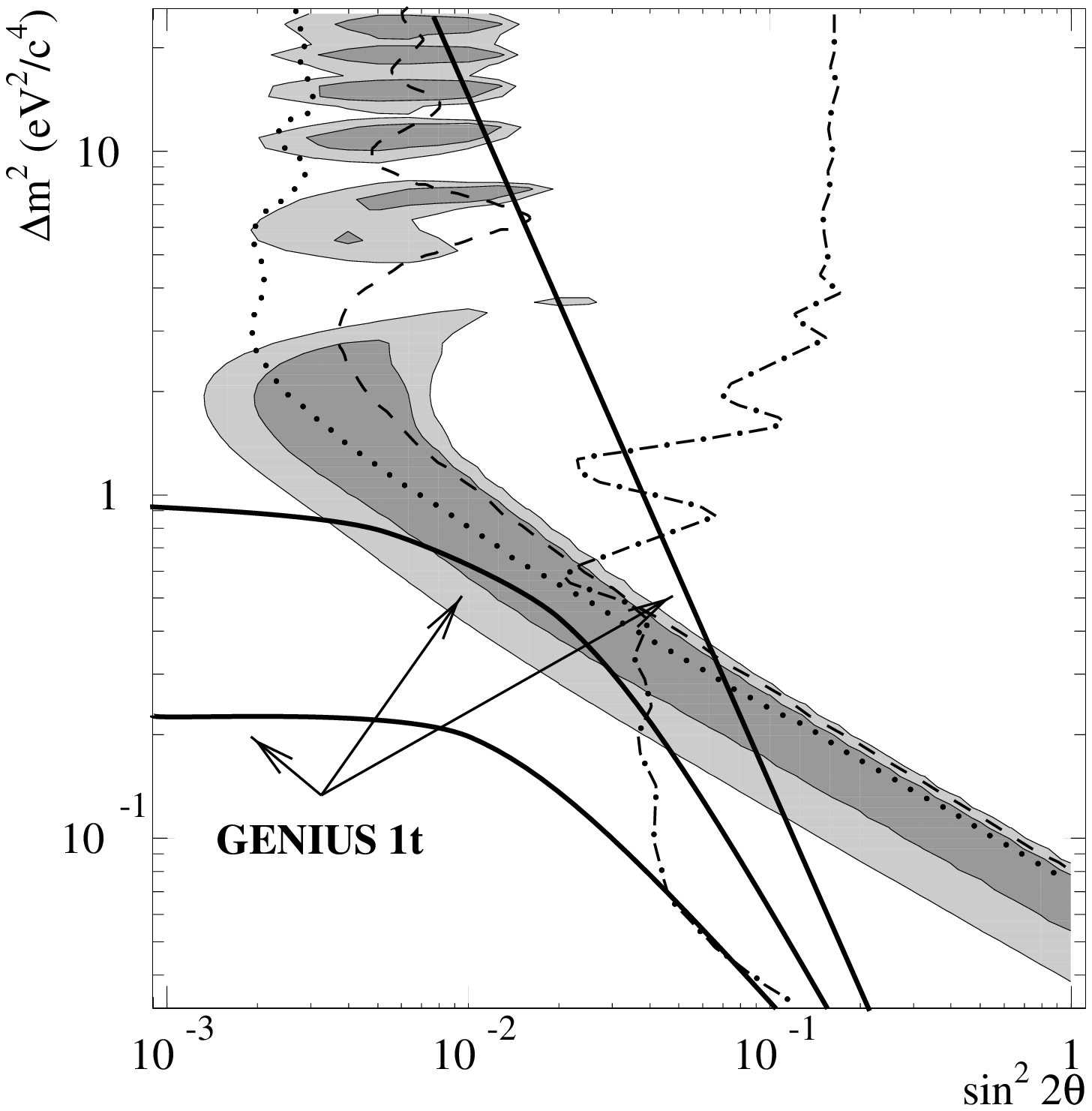}

{\bf Fig. 9}
{\it
LSND compared to the sensitivity of GENIUS 1t 
for $\eta^{CP} = +1$ and three ratios $R_{12}$, from top to bottom 
$R_{12}= 0, 0.01, 0.02$ (from [Kla97c]}
\end{figure}
Already in the worst case for double beta decay of $m_1/m_2=0$
GENIUS 1 ton is more sensitive than the running CERN experiments.
For quasi--degenerate models, for example $R=0.01$ already, GENIUS 1 ton would 
be more sensitive than the planned future accelerator neutrino experiments.

Fig. 9 shows the potential of GENIUS for checking the LSND indication for
neutrino oscillations (original figure from \cite{Ath96}). 
Under the assumption 
$m_1/m_2 \geq 0.02$ and $\eta=1$, GENIUS 1 ton will be sufficient to find 
$0\nu\beta\beta$ decay if the LSND result is to be explained in terms of $\nu_e
\leftrightarrow \nu_{\mu}$ oscillations. This might be of particular interest 
also since the upgraded KARMEN will not completely cover \cite{Dre97} the full
allowed LSND range. Fig. 10 shows a summary of currently known constraints
on neutrino oscillation parameters (original taken from \cite{Hat94}), but 
including the $0\nu\beta\beta$ decay sensitivities of GENIUS 1 ton and GENIUS 
10 tons, for different assumptions on $m_1/m_2$ (for $\eta^{CP}=+1$,
for $\eta^{CP}=-1$ see \cite{Kla97d}).
It is seen that already GENIUS 1 ton tests all degenerate or quasi--degenerate
($m_1/m_2 \geq \sim 0.01$) 
neutrino mass models in any range where neutrinos are 
interesting for cosmology, and also the atmospheric neutrino problem, if it is 
due to $\nu_e \leftrightarrow \nu_{\mu}$ oscillations. GENIUS in its 10 ton
version would directly test the large angle solution of the solar neutrino 
problem and in case of almost degenerate neutrino masses, also the
small angle solution. 

For further recent discussions of the potential of GENIUS for probing neutrino 
mass textures we refer, e.g., to \cite{Kla99b,Giu99,Cza99,Vis99,Bil99}. 

\subsubsection*{GENIUS and super--heavy left--handed neutrinos:}
Fig. 11 (from \cite{Bel98}) compares the sensitivity of GENIUS for heavy 
left-handed neutrinos (as function of $U_{ei}^2$, for which the present
LEP limit is $U_{ei}^2 \leq 5 \cdot 10^{-3}$ \cite{Nar95}) with the discovery 
limit for $e^- e^- \rightarrow W^- W^-$  at Next Linear Colliders. The 
observable in $0\nu\beta\beta$ decay is 
\be
\langle m^{-1}_{\nu} \rangle_H = \sum_i^{''} U^2_{ei} \frac{1}{M_i}. 
\ee
Also shown are the present limits from the Heidelberg--Moscow experiment
(denoted by $0\nu\beta\beta$) assuming different matrix elements. It is
obvious that $0\nu\beta\beta$ is more sensitive than any reasonable future
Linear Collider.

\subsubsection*{GENIUS and left--right symmetry:}
If GENIUS is able to reach down to $\emass \le 0.01$ eV, it would at 
the same time be sensitive to right-handed $W$-boson masses up to 
$m_{W_R} \ge 8$ TeV (for a heavy right-handed neutrino mass of 
$1$ TeV) or $m_{W_R} \ge 5.3$ TeV (at $\langle m_N \rangle = m_{W_R}$) 
\cite{Kla97d}. 
Such a limit would be comparable to the one expected for LHC, 
see for example \cite{Riz96}, which quotes a final sensitivity 
of something like $5-6$ TeV. Note, however that in order to 
obtain such a limit the experiments at LHC need to accumulate 
about $100 fb^{-1}$ of statistics. A 10 ton version of
 GENIUS 
could even reach a sensitivity of $m_{W_R} \ge 18$ TeV (for a heavy 
right-handed neutrino mass of
$1$ TeV) or 
$m_{W_R} \ge 10.1$ TeV (at $\langle m_N \rangle = m_{W_R}$).

This means that already GENIUS 1 ton could be sufficient to definitely
test recent supersymmetric left--right symmetric models having the 
nice features of solving the strong CP problem without the need for an axion 
and having automatic R--parity conservation \cite{Kuc95,Moh96}.

\subsubsection*{GENIUS and $R_p$--violating SUSY:}
The improvement on the R--parity breaking Yukawa coupling $\lambda^{'}_{111}$
(see section 2.2) is shown in Fig. 12.
The full line to the right is the expected sensitivity of the 
LHC -- in the 
limit of large statistics. The three dashed--dotted lines denote (from top
to bottom) the current constraint from the Heidelberg--Moscow experiment
and the sensitivity of GENIUS 1 ton and GENIUS 10 tons, all
 for the 
conservative case of a gluino mass of 1 TeV. If squarks would be heavier than 
1 TeV, LHC could not compete with GENIUS. However, for typical squark masses  
below 1 TeV, LHC could probe smaller couplings.
However, one should keep in 
mind, that LHC can probe squark masses up to 1 TeV only with several years of 
data taking. 

\subsubsection*{GENIUS and $R_p$--conserving SUSY:}
Since the limits on a `Majorana--like' sneutrino mass $\tilde{m}_M$ scale
with $(T_{1/2})^{1/4}$, GENIUS 1 ton (or 10 tons)
would test `Majorana' sneutrino masses lower
by factors of about 7(20), compared with present constraints 
\cite{Hir97,Hir97a,Hir97b}. 

\subsubsection*{GENIUS and Leptoquarks:}
Limits on the lepton--number violating parameters defined in sections
2.7, 3.2
improve as $\sqrt{T_{1/2}}$. This means that for leptoquarks in the range
of 200 GeV LQ--Higgs couplings down to (a few) $10^{-8}$ could be explored. 
In other words, if leptoquarks interact with the standard model Higgs boson
with a coupling of the order ${\cal O}(1)$, either $0\nu\beta\beta$ must be 
found, or LQs must be heavier than (several) 10 TeV. 

\newpage

\vspace*{-20mm} 
\hspace*{-10mm}
\epsfysize=150mm
\epsfbox{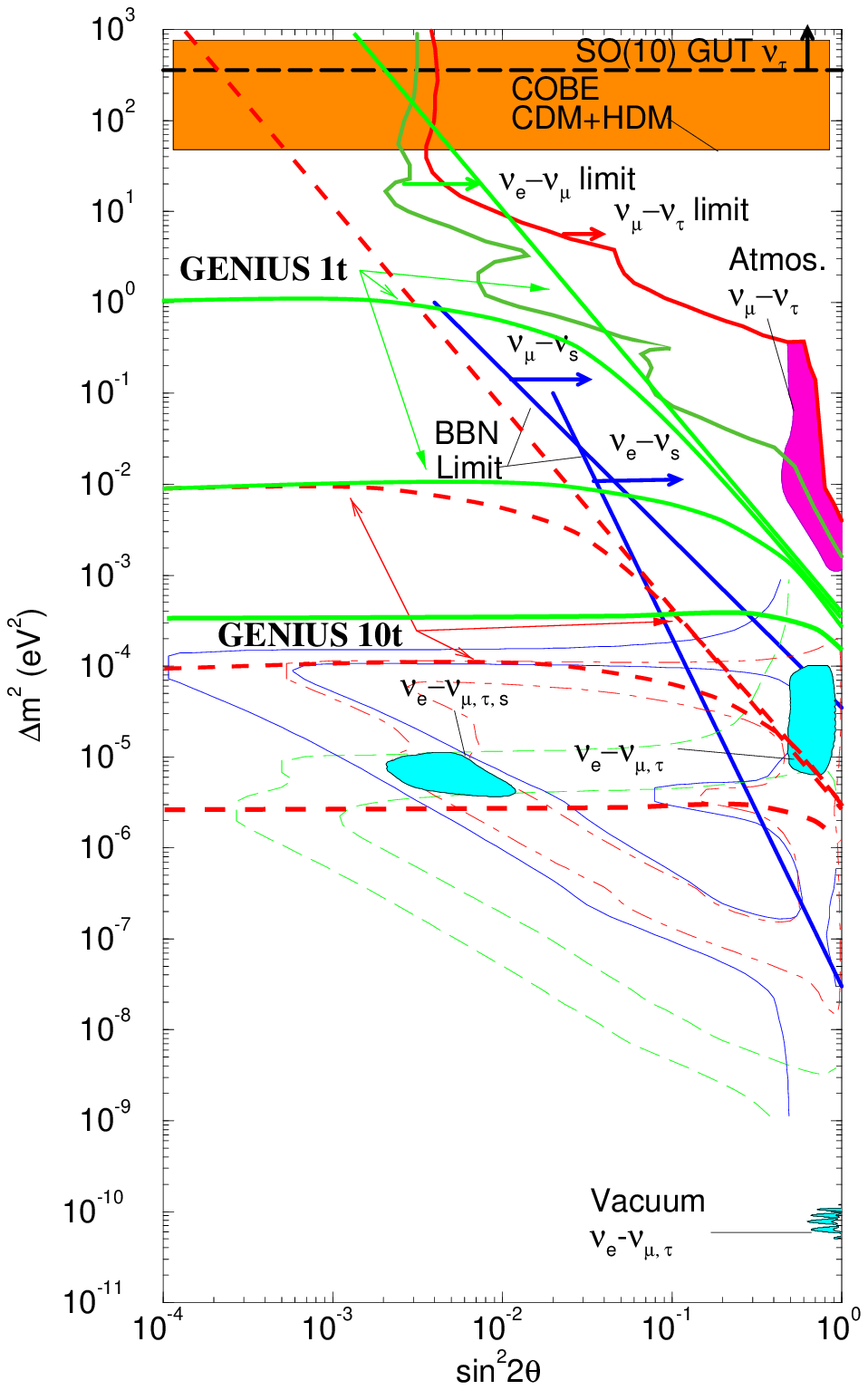}

\noindent
{\bf Fig. 10}
{\it Summary of currently known constraints on neutrino 
oscillation parameters. The (background) figure without the \znbb{} 
decay constraints can be obtained from\\ 
http://dept.physics.upenn.edu/~www/neutrino/solar.html. Shown are 
the vacuum and MSW solutions (for two generations of neutrinos) 
for the solar neutrino problem, 
the parameter range which would solve the atmospheric neutrino problem 
and various reactor and accelerator limits on neutrino oscillations. 
In addition, the mass range in which neutrinos are good hot dark matter 
candidates is indicated, 
as well as limits on neutrino oscillations into sterile states from 
considerations of big bang nucleosynthesis. Finally the 
thick lines indicate the sensitivity of GENIUS (full lines 1 ton, 
broken lines 10 ton) to neutrino oscillation parameters for three values 
of neutrino mass ratios $R = 0, 0.01$ and $0.1$ (from top to bottom).
For GENIUS 10 ton also the contour line for $R=0.5$ is shown.  
The region beyond the lines would be excluded.
While already the 1 ton GENIUS would be sufficient to constrain degenerate 
and quasi-degenerate neutrino mass models, and also would solve the 
atmospheric neutrino problem if it is due to 
$\nu_e \leftrightarrow \nu_{\mu}$ oscillations, the 10 ton version of 
GENIUS could cover a significant new part of the parameter space, 
including the large angle MSW solution to the solar neutrino problem, 
even in the worst case of $R=0$. For $R\geq 0.5$ it would even probe the 
small angle MSW solution (see \cite{klapneut,KKP}).} 
\bigskip

\subsubsection*{GENIUS and composite neutrinos}
GENIUS in the 1(10) ton version would improve the limit on the excited
Majorana neutrino mass deduced from the Heidelberg--Moscow experiment
(eq. 32) to
\be
m_N\geq  1.1 (2.3) \hskip3mm TeV
\ee

\begin{figure}[!h]
\hspace*{-1cm}
\epsfysize=60mm\centerline{\epsfbox{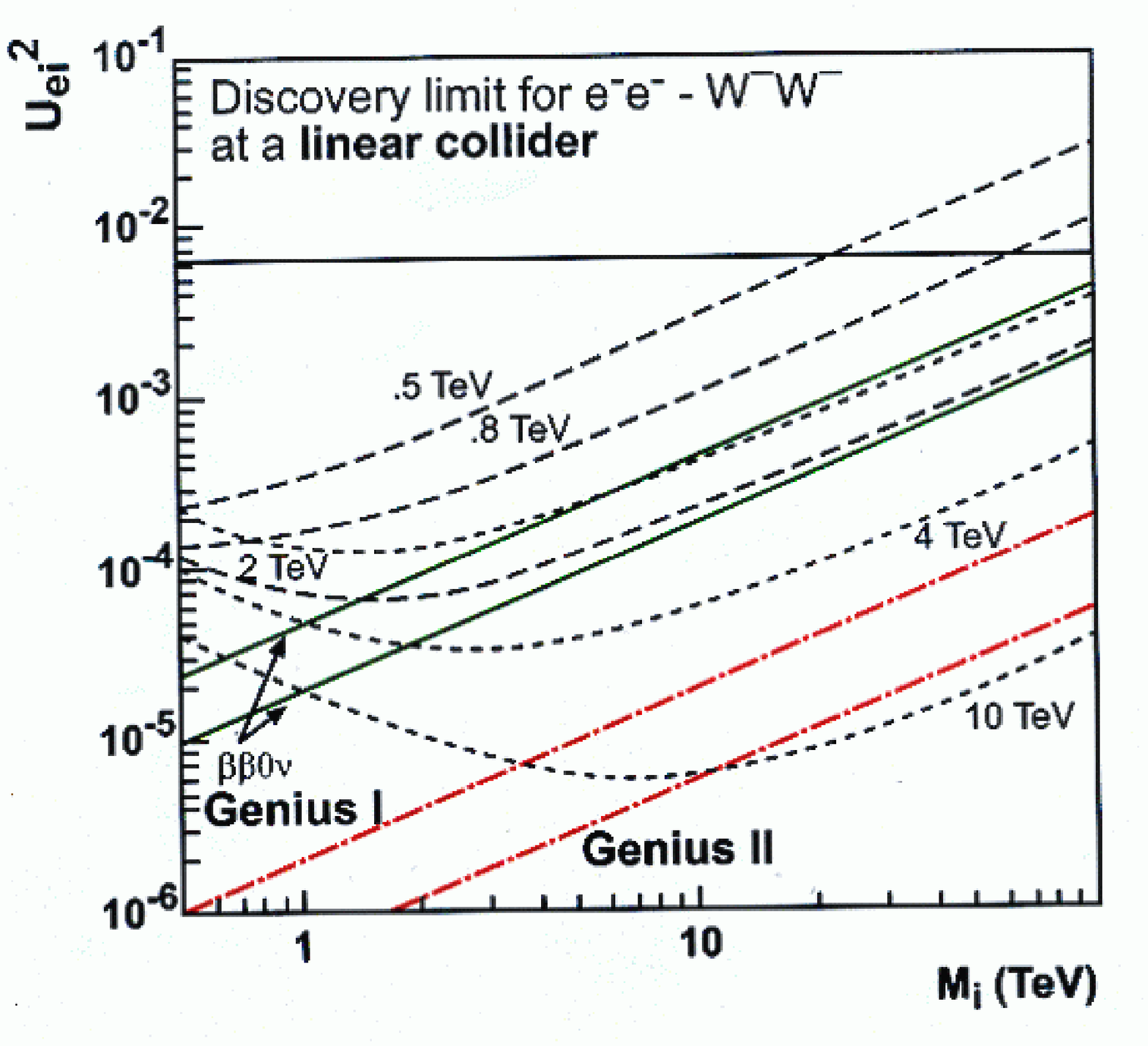}}
\\
{\bf Fig. 11}
{\it Discovery limit for $e^- e^- \rightarrow W^- W^-$ at a linear collider 
as function of the mass $M_i$ of a heavy left--handed neutrino, and of
$U_{ei}^2$ for $\sqrt{s}$ between 500 GeV and 10 TeV. In all cases the
parameter space above the line corresponds to observable events. 
Also shown are the limits set by the Heidelberg--Moscow $0\nu\beta\beta$
experiment as well as the prospective limits from GENIUS. The areas {\rm above}
the $0\nu\beta\beta$ contour lines are {\rm excluded}. The horizontal
line denotes the limit on neutrino mixing, $U_{ei}^2$, from LEP.
Here the parameter space above the line is excluded. (from \cite{Bel98}).
}
\end{figure}

A recent detailed study \cite{Pan99} shows that while the HEIDELBERG--MOSCOW
experiment already exceeds the sensitivity of LEPII in probing compositeness,
GENIUS will reach the sensitivity of LHC. With the $0\nu\beta\beta$ half life
against decay by exchange of a composite Majorana 
neutrino given by \cite{Pan99}
\be
T_{1/2}^{-1}=\Big(\frac{f}{\Lambda_c}\Big)^4 \frac{m_A^8}{M_N^2} 
|{\cal M}_{FI}|^2 \frac{G_{01}}{m_e^2}
\ee
where $M_N$ is the composite neutrino Majorana mass, and $f$ denotes the 
coupling with the electron, figure 13 shows the situations of GENIUS and LHC. 

\subsubsection{GENIUS, special relativity and equivalence principle
in the neutrino sector\\}

The already now strongest limits given by the Heidelberg--Moscow experiment
discussed in section 3.2 would be improved by 1--2 orders of magnitude.
It should be stressed again, that while neutrino oscillation bounds 
constrain the region of large mixing of the weak and gravitational 
eigenstates, these bounds from double beta decay apply even in the case
of no mixing and thus probe a totally unconstrained region in the parameter 
space.

\begin{figure}[t]
\vskip-25mm 
\hskip10mm
\epsfxsize=100mm
\epsfysize=120mm
\epsfbox{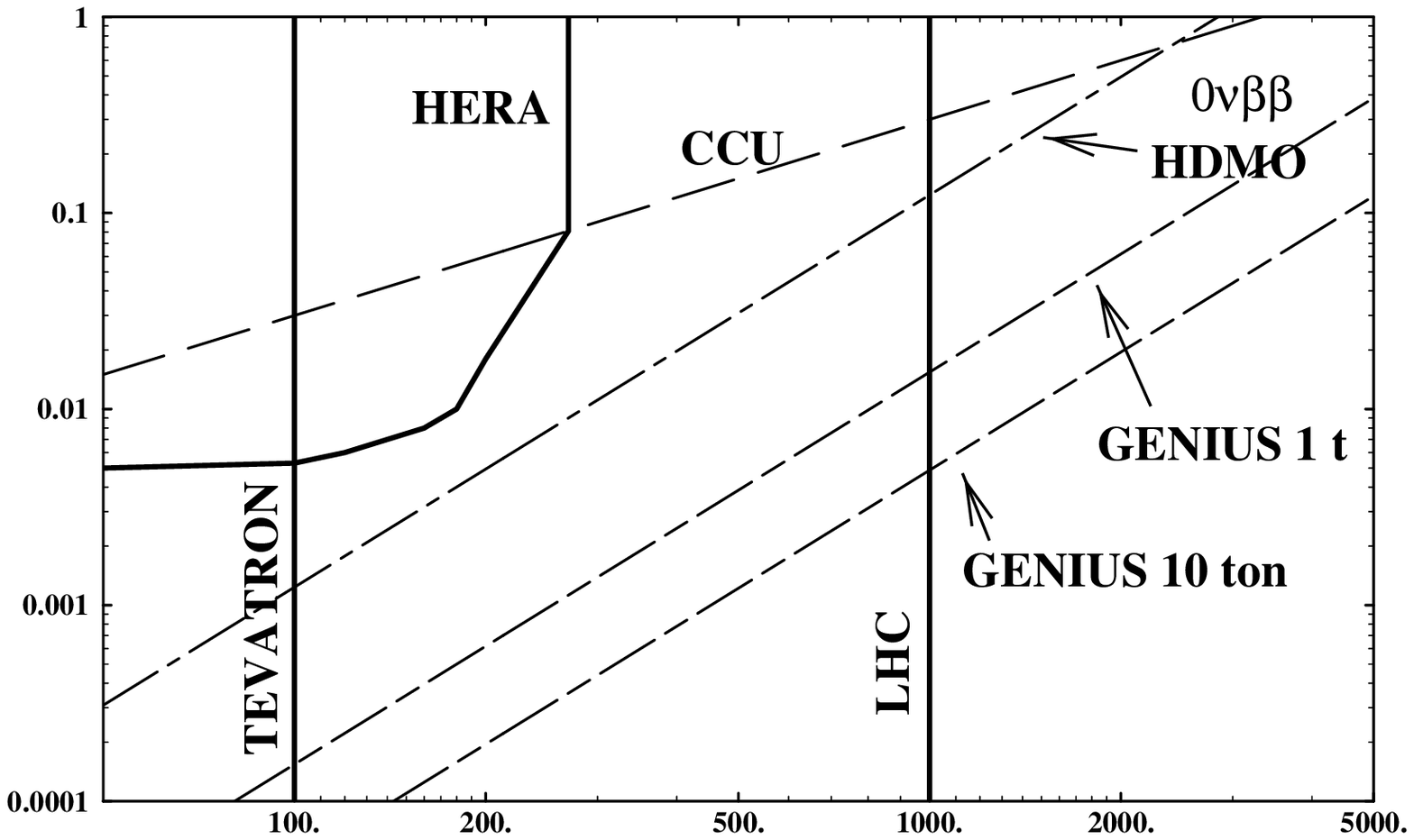}

\vskip-80mm
\noindent
$\lambda'_{111}$ 

\vskip45mm 
\hskip90mm $m_{\tilde q}$ [GeV] 

\bigskip

{\bf Fig. 12}
{\it Comparison of sensitivities of existing and future 
experiments on \rp SUSY models in the plane $\lambda'_{111}-m_{\tilde q}$. 
Note the double logarithmic scale! Shown are the areas currently excluded 
by the experiments at the TEVATRON, the limit from charged-current 
universality, denoted by CCU, and the limit from absence of \znbb{} 
decay from the Heidelberg-Moscow collaboration (\znbb{} HDMO). 
In addition, the estimated sensitivity of HERA and the LHC is compared to the 
one expected for GENIUS in the 1 ton and the 10 ton version.}
\end{figure}

\begin{figure}[t]
\epsfysize=100mm
\centerline{\epsfbox{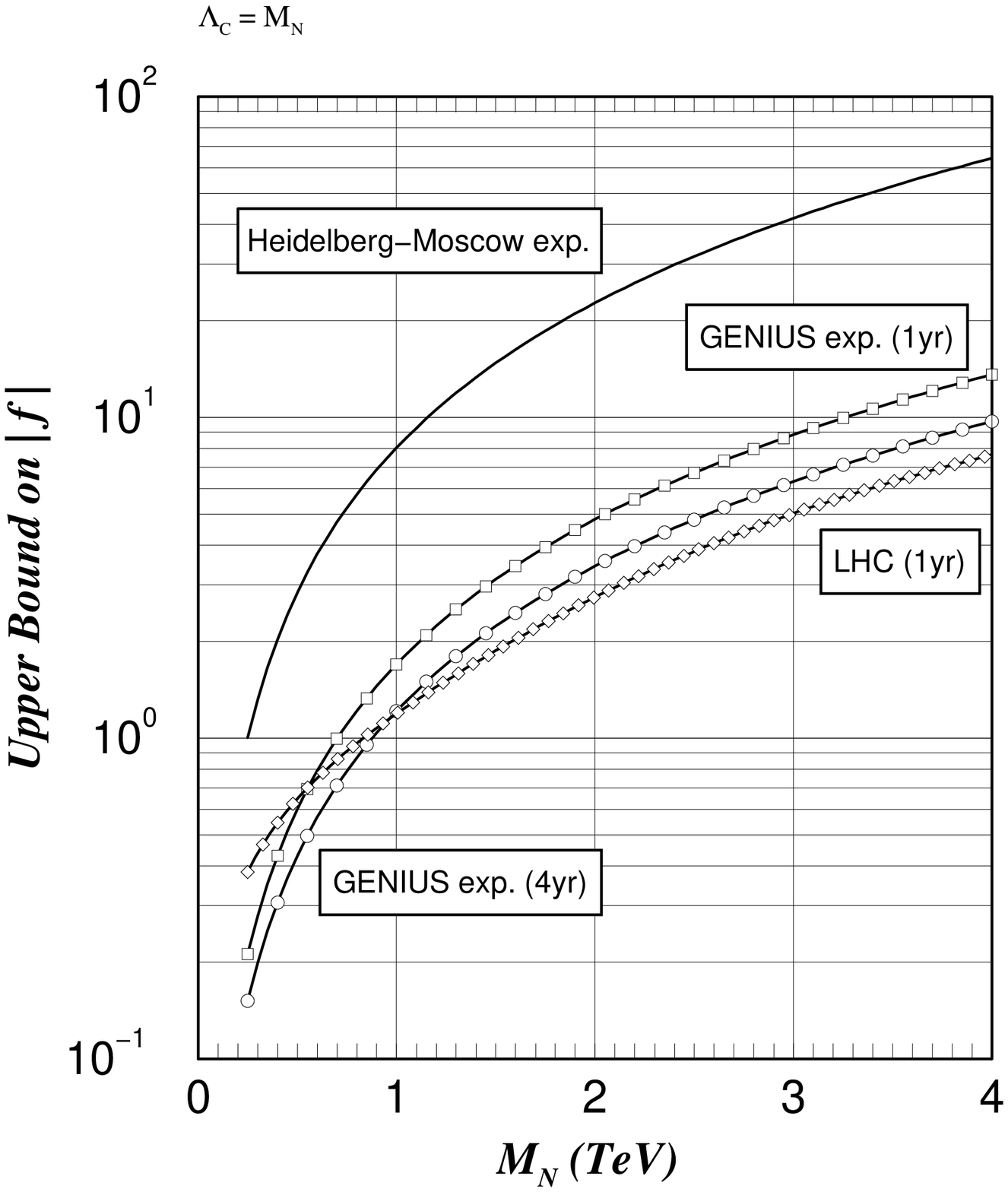}}
{\bf Fig. 13}
{\it Sensitivity of LHC and GENIUS to compositeness parameters (assuming 
$\Lambda_C=M_N$). Regions above the curves are excluded. The LHC bound is weaker
than the GENIUS bound for $M_N<550 (1000)$ GeV. 
(from \cite{Pan99}
}
\end{figure}

\begin{figure}[!h]
\hspace*{-1cm}
\epsfysize=60mm\centerline{\epsfbox{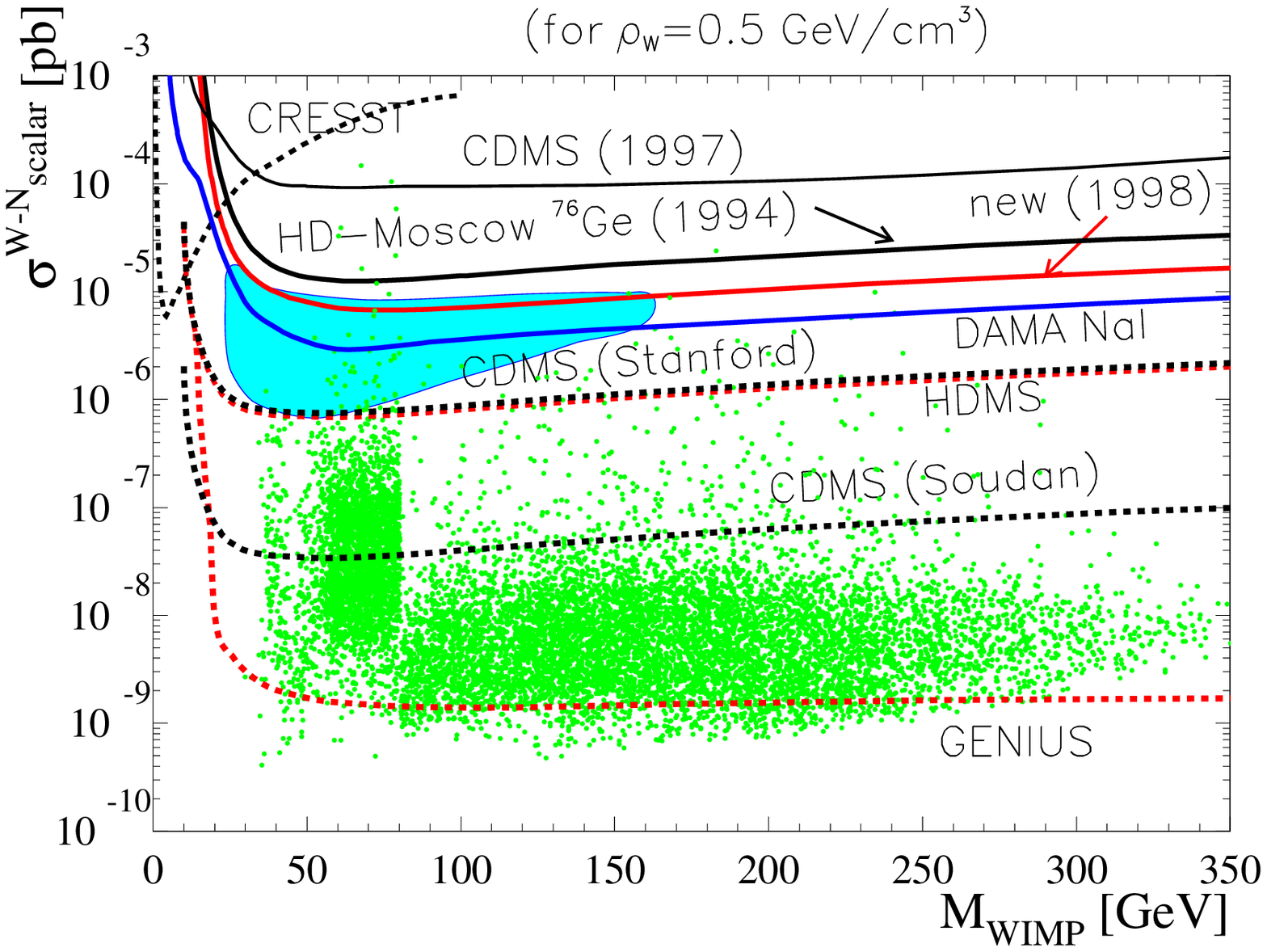}}
\\
{\bf Fig. 14}
{\it WIMP--nucleon cross section
limits in pb for scalar interactions as
function of the WIMP--mass in 
GeV. 
Regions beyond solid lines are excluded by 
experiment \cite{101,HM98,Ber97,Ake97}.
Further shown are expected sensitivities of experiments under construction
 (dashed lines for
HDMS \protect{\cite{Bau97,Kla97e}}, CDMS \protect{\cite{Ake97}}, CRESST and for
GENIUS). These 
limits are compared to theoretical
expectations (scatter plot) for WIMP--neutralino cross sections calculated 
in the
MSSM framework with non--universal scalar mass unification
\protect{\cite{Bed97b}}.  
The 90~\% allowed region claimed by \protect{\cite{Ber97a}}
(light filled area), which is further
restricted by indirect dark matter searches \protect{\cite{Bot97}} 
(dark filled
area), could already be easily tested with a 100 kg version of the GENIUS 
experiment.}
\end{figure}

\subsubsection{GENIUS and dark matter\\}

{\it Neutrinos as hot dark matter}\\
If neutrinos have masses in the range of a few eV, they would be good 
candidates for the hot dark matter in the universe. From the dark 
matter argument itself it does not follow which neutrino has to be in this mass
range. Clearly, if a neutrino with sizeable mixing angle to the electron 
neutrino in this mass range exists, one expects GENIUS to find 
$0\nu\beta\beta$ decay.

However, if the $\nu_{\tau}$ is in the eV range, the $\nu_e$ and $\nu_{\mu}$
being lighter by at least factors of hundreds and the the 
$\nu_{\tau}-\nu_e$ mixing angle small at the same time GENIUS with 1 ton 
would not find double beta decay. In the case of quasidegenerate models
or degenerate models, on the other hand, $0\nu\beta\beta$ decay should be found
by GENIUS, unless the CP--phases between the different mass eigenstates 
take on some special combinations and have a relative minus sign, see the
discussion in \cite{KK3}.\\
{\it Cold Dark Matter}\\
Weakly interacting massive particles (WIMPs) are candidates for the 
cold dark matter in the universe. The favorite WIMP candidate is 
the lightest supersymmetric particle, presumably the neutralino. 
The expected detection rates for neutralinos of typically less 
than one event per day and kg of detector mass
\cite{Bed94,Bed97a,Bed97b,Jun96}, 
however, make direct searches for WIMP scattering experimentally 
a formidable task. 

Fig. 14 shows a comparison of existing constraints and future 
sensitivities of cold dark matter experiments, together with the 
theoretical expectations for neutralino scattering rates 
\cite{Bed97b}.
Obviously, GENIUS could easily cover the range of positive 
evidence for dark matter
recently claimed by DAMA \cite{Ber97a,Bot97}. 
It would also be by far more
sensitive than all other dark matter experiments at present under construction
or proposed, like the cryogenic experiment CDMS. Furthermore,
obviously GENIUS will be the only experiment, 
which could seriously test the MSSM predictions over the whole 
SUSY parameter space. In this way, GENIUS could compete even 
with LHC in the search for SUSY, see for example the discussion 
in \cite{Bae97}. It is important to note, that GENIUS could reach the 
sensitivity shown in Fig. 14 with only 100 kg of {\it natural} Ge
detectors in a measuring time of three years \cite{Kla98d}.  

 Finding the neutralino with GENIUS would imply typical 
limits on R-parity violating couplings of the order of $10^{-(16-20)}$ 
for any of the $\lambda_{ijk}$, $\lambda'_{ijk}$ or $\lambda''_{ijk}$ 
in the superpotential (eq. 11). \\

{\it 4.2.5 GENIUS and solar neutrinos}\\
The potential of GENIUS to measure the spectrum of low energy solar neutrinos 
in real time has been studied by \cite{Bau99b}. The detection reaction is
elastic neutrino electron scattering, $\nu+e \rightarrow \nu+ e$. 
The energy threshold is a few keV, the expected number of events for a
target
of one ton of (natural or enriched)
Germanium is 3.6 events/day in the standard solar model. 
Achieving a background low enough to measure the low energy solar neutrino 
spectrum should be possible.

\section{Conclusion}
Double beta decay has a broad potential for providing important information
on modern particle physics beyond present and future high energy accelerator
energies which will be competitive for the next decade and more. 
This includes SUSY 
models, compositeness, left--right symmetric models, leptoquarks,  the
neutrino and sneutrino mass and tests of Lorentz invariance and equivalence 
principle in the neutrino sector. 
Results have been deduced from the HEIDELBERG--MOSCOW
experiment for these topics and have been presented.
For the neutrino mass double beta decay now is particularly 
pushed into a key position by the recent possible indications of beyond 
standard model physics from the side of solar and atmospheric neutrinos,
dark matter COBE results and others. 
Neutrino mass scenarios which could explain these observations,
can be checked already now by double beta decay. The HEIDELBERG--MOSCOW 
experiment has reached a leading position among present $\beta\beta$
experiments and as the first of them yields results in the sub--eV
range -- with striking consequences on presently discussed neutrino mass 
textures. 
 
A future double beta experiment
(GENIUS) with highly increased sensitivity based on use of 1 ton or more
of enriched `naked' $^{76}$Ge detectors in liquid nitrogen
would be a breakthrough into the multi-TeV range for many 
beyond standard models. The sensitivity for the neutrino mass would reach
down to 0.01 or even 0.001 eV. The experiment would be competitive to LHC with 
respect to the mass of a right--handed W boson, in search for R--parity 
violation and others, and would improve the leptoquark and compositeness 
searches
by considerable factors. It would probe the Majorana electron
sneutrino mass more
sensitive than NLC (Next Linear Collider). It would yield constraints on 
neutrino oscillation parameters far beyond all present terrestrial 
$\nu_e - \nu_x$ neutrino oscillation experiments and could test directly the
large and, for degenerate models,
even the small angle solution of the solar neutrino
problem. GENIUS would cover the full SUSY parameter space for prediction of 
neutralinos as cold dark matter and compete in this way with LHC in the search
for supersymmetry. Even if SUSY would be first observed by LHC, it would
still be fascinating to verify the existence and properties of neutralino
dark matter, which could be achieved by GENIUS. 
GENIUS could also serve as a first real time detector for solar pp-neutrinos.
Concluding, GENIUS
has the ability to provide a major tool for future particle-- and astrophysics.
    
 Finally it may be stressed that the technology of producing and using
enriched high purity germanium detectors, which have been produced for the 
first time for the Heidelberg--Moscow experiment, has found 
meanwhile applications also in pre-GENIUS dark matter search
\cite{101,102,Kla97e,Bau97} and in high--resolution $\gamma$-ray astrophysics,
using balloons and satellites \cite{Kla91,81,109}
\cite{100,111,Kla97b,Kla98h}.


\begin{thebibliography}{200}






\bibitem[Adh98]{Adh98} 
R. Adhikari, G. Rajasekaran, hep-ph/9812361vs3, Dec1998

\bibitem[Ake97]{Ake97}
D.S. Akerib et al. (CDMS collab.), preprint astro-ph/9712343






\bibitem[Ale94]{91} A. Alessandrello et al., Phys. Lett {\bf B 335} (1994) 519
 



\bibitem[Als89]{89} M. Alston-Garnjost et al., Phys. Rev Lett. {\bf 7}1 (1993)
 831

\bibitem[Alt97]{Alt97}
G. Altarelli, J. Ellis, G.F. Guidice, S. Lola, M.L. Mangano,
preprint hep-ph/9703276


\bibitem[Ath96]{Ath96}
C. Athanassopoulos {\it et al.}, LSND collab., Phys. Rev. {\bf C 54}
(1996) 2685, Phys. Rev. Lett. {\bf 77} (1996) 3082 



\bibitem[Bab95]{bab95}
K.S. Babu, R.N. Mohapatra, Phys. Rev. Lett. {\bf 75} (1995) 2276

\bibitem[Bab97a]{Bab97}
K.S. Babu et al., preprint hep-ph/9703299 (March 1997)

\bibitem[Bab97b]{Bab97b}
K.S. Babu {\it et al.}, preprint hep-ph/9705414v2 (1997)

\bibitem[Bae97]{Bae97}
H. Baer. M. Bhrlik, hep-ph/9706509

\bibitem[Bah98]{Bah98}
J.N. Bahcall, P.I. Krastev, A.Yu. Smirnov, Phys. Rev. {\bf D 58}
(1998) 096016

\bibitem[Bam95a]{77}
P. Bamert, C.P. Burgess, R.N. Mohapatra, Nucl. Phys. {\bf B 438} (1995) 3


\bibitem[Bam95b]{69}   
P. Bamert, C.P. Burgess, R.N. Mohapatra, Nucl. Phys. {\bf B 449} (1995)
25 





\bibitem[Bar93]{109}
S.D. Barthelmy {\it et al.}, AIP Conf. Proc. 280 (1993) 1166




\bibitem[Bar94]{100} S. D. Barthelmy et al., Astrophys. 
J. {\bf 427} (1994) 519 


   
\bibitem[Bar97]{Bar97}
A.S. Barabash, Proc. NEUTRINO 96, Helsinki, June 1996, World Scientific, 
Singapore (1997), p. 374

\bibitem[Bar98]{Bar98}
G. Barenboim, F. Scheck, hep-ph/9808327

\bibitem[Bau97]{Bau97}
L. Baudis, J. Hellmig, H.V. Klapdor--Kleingrothaus, A. M\"uller, F. Petry,
Y. Ramachers, H. Strecker, Nucl. Inst. Meth.  {\bf A 385} (1997) 265

\bibitem[Bau98]{Bau98}
 L.~Baudis, G.~Heusser, B.~Majorovits, Y.~Ramachers, H.~Strecker,
 H.V.~Klapdor--Kleingrothaus, hep-ex/9811040 and Nucl. Instr. Meth. 
{\bf A 426} (1999) 425

\bibitem[Bau99a]{Bau99a}
 L.~Baudis et al. (Heidelberg--Moscow collab.), hep-ex/9902014 and 
Phys. Rev. {\bf D}, in press (1999)
 
\bibitem[Bau99b]{Bau99b}
L. Baudis, H.V. Klapdor--Kleingrothaus, Eur. Phys. J. A, in press 
and hep-ex/9906044 


\bibitem[Bed94]{Bed94}
V. Bednyakov, H.V. Klapdor--Kleingrothaus, S.G. Kovalenko, Phys. Lett. 
{\bf B 329} (1994) 5 and Phys. Rev. {\bf D 50} (1994) 7128

\bibitem[Bed97a]{Bed97a}
V. Bednyakov, H.V. Klapdor--Kleingrothaus, S.G. Kovalenko, 
Phys. Rev. {\bf D 55} (1997) 503

\bibitem[Bed97b]{Bed97b}
V. Bednyakov, H.V. Klapdor--Kleingrothaus, S.G. Kovalenko, 
Y. Ramachers,
Z. Phys. {\bf A 357} (1997) 339

\bibitem[Bed97c]{Bed97c}
V. Bednyakov, V.B. Brudanin, S.G. Kovalenko, Ts. D. Vylov, Mod. Phys. Lett 
{\bf A 12} (1997) 233

\bibitem[Bed98]{Bed98}
V. Bednyakov, A. Faessler, S. Kovalenko, hep-ph/9808224v2


\bibitem[Bel96]{14}
G. Belanger, F. Boudjema, D.London, H. Nadeau, Phys. Rev. {\bf D 53} (1996) 
6292

\bibitem[Bel98]{Bel98}
G. Belanger, Proc. Int. Conf. on Lepton and Baryon Number Violation in 
Part. Phys. and Cosmology, Trento, Italy, april 20--25, 1998, editors: 
H.V.~Klapdor-Kleingrothaus and I.~Krivoshina (IoP, Bristol, 1999)



\bibitem[Ber95]{Ber95}
Z. Berezhiani, R.N. Mohapatra, Phys. Rev. {\bf D 52} (1995) 6607

\bibitem[Ber97a]{Ber97}
R. Bernabei {\it et al.}, Phys. Lett. {\bf B 389} (1997) 757


\bibitem[Ber97b]{Ber97a}
R. Bernabei {\it et al.}, ROM2F/97/33; P. Belli at TAUP, Gran Sasso, Sept. 7 
(1997)

\bibitem[Bha97]{Bha97}
G. Bhattacharyya, in \cite{Kla98}


\bibitem[Bil99]{Bil99}
S.M. Bilenky {\it et al.}, hep-ph/9907234


\bibitem[Boc94]{111}
J. Bockholt, H.V. Klapdor--Kleingrothaus, Nucl. Phys. {\bf B}
(Proc. Suppl.) 35 (1994) 403

\bibitem[Bot97]{Bot97}
A. Bottino {\it et al.}, hep-ph/9709292

\bibitem[Buc87]{Lagr1} W. Buchm\"uller, R. R\"uckl and D. Wyler,
                       Phys.Lett. B191 (1987) 442.




\bibitem[Bur93]{68}
C.P. Burgess, J.M. Cline, Phys. Lett. {\bf B 298} (1993) 141; 
Phys. Rev. {\bf D 49} (1994) 5925

\bibitem[Bur96]{72}
C.P.Burgess, in \cite{tren}

\bibitem[Bur99]{Bur99}
S. Burles, K.M. Nollet, J.N. Truram, M.S. Turner, astro-ph/9901157

\bibitem[But93]{nu3}M.\ N.\ Butler {\it et al.}, Phys.\ Rev.\ {\bf D47},
  2615 (1993); A.\ Halprin and C.\ N.\ Leung, Phys.\ Rev.\ Lett. {\bf 67},
  1833 (1991); J.\ Pantaleone, A.\ Halprin, and C.\ N.\ Leung, Phys.\ Rev.\
  {\bf D47}, R4199 (1993); K. Iida, H. Minakata and O. Yasuda,
  Mod. Phys. Lett. {\bf A8} (1993) 1037.






\bibitem[Cal93]{Cal93}
D.O. Caldwell, R.N. Mohapatra, Phys. Rev. {\bf D 48} (1993) 3259


\bibitem[Car93]{70}
C.D. Carone, Phys. Lett. {\bf B 308} (1993) 85









\bibitem[Cho97]{Cho97}
D. Choudhury, S. Raychaudhuri, preprint hep-ph/9702392

\bibitem[Cli97]{CLI96} D. Cline, in: Proceedings of the {\it International 
Workshop Dark Matter in Astro-- and Particle Physics} (DARK96),
Eds. H.V. Klapdor--Kleingrothaus, Y. Ramachers, World Scientific 1997,
p. 479

\bibitem[Col97]{cole} S. Coleman and S.L. Glashow, Phys. Lett. {\bf B 405},
  249 (1997).

\bibitem[Cza99]{Cza99}
M. Czakon, J. Gluza, M. Zralek, hep-ph/9906381

\bibitem[Dan95]{90} F. A. Danevich et al., Phys. Lett. {\bf B 344} (1995) 72









\bibitem[Doi85]{74}
M. Doi, T. Kotani, E.Takasugi, Progr. Theor. Phys. Suppl. {\bf 83} (1985) 1



\bibitem[Doi93]{49}
M. Doi, T. Kotani, Progr. Theor. Phys. 89 (1993) 139

\bibitem[Dre97]{Dre97}
G. Drexlin, Proc. Internat. School on Neutrino Physics, Erice, Italy,
Sept. 1997, to be publ. in Plenum Press

\bibitem[Ell87]{Ell87}
S.R. Elliot, A.A. Hahn, M.K. Moe, Phys. Rev. Lett. {\bf 59} (1987) 1649

\bibitem[Ell92]{88} S. R. Elliott et al., Phys. Rev. {\bf C 46} (1992) 1535


\bibitem[Fal94]{102}
T. Falk, A. Olive, M. Srednicki, Phys. Lett. {\bf B339} (1994) 248

\bibitem[Fel98]{Fel98}
G. J. Feldman and R. D. Cousins, Phys. Rev. {\bf D57} (1998) 3873 

\bibitem[Fri88]{62}
J.Friemann, H.Haber, K.Freese, Phys. Lett. {\bf B 200} (1988) 115;
J. Bahcall, S. Petcov, S. Toshev and J.W.F. Valle, Phys.Lett. {\bf B 181}
(1986) 369; Z. Berezhiani and M. Vysotsky, Phys. Lett. {\bf B 199}
(1988) 281.




\bibitem[Fri95]{22}
H. Fritzsch and Zhi-zhong Xing, preprint hep-ph/9509389,
Phys. Lett. {\bf B 372} (1996) 265

\bibitem[Fuk99]{Fuk99}
Y. Fukuda {\it et al.}, Phys. Rev. Lett. 82 (1999) 1810   

\bibitem[Gas89a]{nu1}M. Gasperini, Phys. Rev. {\bf D38}, 2635 (1988); {\it
  ibid.} {\bf D39}, 3606 (1989).

\bibitem[Gas89b]{gasp}M. Gasperini, Phys. Rev. Lett. {\bf 62}, 1945 (1989).




\bibitem[Geo81]{61}  
H.M. Georgi, S.L.Glashow and S. Nussinov, Nuc. Phys. {\bf B 193} (1981)
297



\bibitem[Giu99]{Giu99}
C. Giunti, hep-ph/9906275 (June 1999)

\bibitem[Gla97]{glash} S.L. Glashow, A. Halprin, P.I. Krastev, C.N.
  Leung, and J. Panteleone, Phys. Rev. {\bf D 56}, 2433 (1997).


\bibitem[Gon95]{Gon95}
M. Gonzalez--Garcia, hep-ph/9510419

\bibitem[Goo61]{vepk} M.L. Good, Phys. Rev. {\bf 121}, 311 (1961);
  O. Nachtmann, Acta Physica Austriaca, Supp. VI
  {\sl Particle Physics} ed. P. Urban,  (1969) p. 485;
  S.H. Aronson, G.J. Bock, H-Y Cheng and E. Fishbach,
  {\bf 48}, 1306 (1982); Phys. Rev. {\bf D28}, 495 (1983);
  I.R. Kenyon, Phys. Lett. {\bf B237}, 274 (1990);
  R.J. Hughes, Phys. Rev. {\bf D46}, R2283 (1992);







\bibitem[Gro90]{16}
K. Grotz, H.V. Klapdor, 'The Weak Interaction in Nuclear, Particle and 
Astrophysics' (Adam Hilger: Bristol, Philadelphia) 1990




\bibitem[H196]{H196}
H1 Collab., S. Aid et al., Phys. Lett. B 369 (1996) 173

\bibitem[Hab93]{44}
H. E. Haber, in Proc. on Recent Advances in the Superworld, Houston,
April 14-16 (1993), hep-ph/9308209


\bibitem[Hal84]{hal84}
L. Hall, M. Suzuki, Nuclear Physics {\bf B 231} (1984) 419

\bibitem[Hal91]{nu2}A. Halprin and C.N. Leung, Phys. Rev. Lett. {\bf 67}, 1833 
  (1991); Nucl. Phys. {\bf B28A} (Proc. Supp.), 139 (1992); 
  J.\ N.\ Bahcall, P.\ I.\ Krastev, and C.\ N.\ Leung, 
  Phys. Rev. {\bf D 52}, 1770 (1995);
  R.B. Mann and U. Sarkar, Phys. Rev. Lett {\bf 76} (1996) 865;
\bibitem[Hal96]{hal}
A. Halprin, C.N. Leung, J. Pantalone,  Phys.Rev. {\bf D 53} (1996) 5365


\bibitem[Ham98]{hambye}
T. Hambye, R.B. Mann and U. Sarkar, Phys. Lett. {\bf B 421},
  105 (1998); Phys. Rev. {\bf D 58}, 025003 (1998).


\bibitem[Hat94]{Hat94}
N. Hata, P. Langacker, Phys. Rev. {\bf D 50} (1994) 632



\bibitem[Hel97]{Hel97}
J. Hellmig, H.V. Klapdor--Kleingrothaus, Z. Phys. {\bf A 359} (1997)
351

\bibitem[Hew97]{Hew97}
J.L. Hewett, T.G. Rizzo, preprint hep-ph/9703337v3 (May1997)


\bibitem[Hir95a]{6}
M. Hirsch, H.V. Klapdor--Kleingrothaus, S.G. Kovalenko, Phys. Rev. Lett. 75
(1995) 17




\bibitem[Hir95b]{47}
M. Hirsch, H.V. Klapdor--Kleingrothaus, S. Kovalenko, Phys. Lett. {\bf B 352}
(1995) 1



\bibitem[Hir96a]{hir96}
M. Hirsch, H.V. Klapdor--Kleingrothaus, S.G. Kovalenko,
Phys. Lett. {\bf B 372} (1996) 181, Erratum: Phys. Lett. {\bf B381} (1996) 488 


\bibitem[Hir96b]{hir96a}
M. Hirsch, H.V. Klapdor--Kleingrothaus, S.G. Kovalenko,
Phys. Lett. {\bf B 378} (1996) 17 and Phys. Rev. {\bf D 54}
(1996) R4207


\bibitem[Hir96c]{75}
M. Hirsch, H. V. Klapdor--Kleingrothaus, S. G. Kovalenko, H. Paes,
Phys. Lett.  {\bf B 372} (1996) 8



\bibitem[Hir96d]{hir96c}
M. Hirsch, H.V. Klapdor--Kleingrothaus, S. Kovalenko, Phys. Rev. {\bf D 53}
(1996) 1329


\bibitem[Hir96e]{11}
M. Hirsch, H.V. Klapdor--Kleingrothaus, in \cite{tren};
M. Hirsch, H.V. Klapdor--Kleingrothaus, O. Panella, 
Phys. Lett. {\bf B 374} (1996) 7

\bibitem[Hir97a]{Hir97}
M. Hirsch, H.V. Klapdor--Kleingrothaus, S.G. Kovalenko, 
Phys. Lett. {\bf B 398} (1997) 311 and {\bf 403}
(1997) 291

\bibitem[Hir97b]{Hir97b}
M. Hirsch, H.V. Klapdor--Kleingrothaus, S. Kovalenko, in \cite{Kla98} 

\bibitem[Hir97c]{Hir97c}
M. Hirsch, H.V. Klapdor--Kleingrothaus, Proc. Int. Workshop on Dark 
Matter
in Astro-- and Particle Physics (DARK96), Heidelberg, Sept. 1996,
Eds. H.V. Klapdor--Kleingrothaus and Y. Ramachers (World Scientific,
Singapore) 1997, p. 640

\bibitem[Hir98a]{Kolb1}
M. Hirsch, H.V. Klapdor--Kleingrothaus, St. Kolb, S.G. Kovalenko, 
Phys. Rev. {\bf D 57} (1998) 2020

\bibitem[Hir98b]{Hir97a}
M. Hirsch, H.V. Klapdor--Kleingrothaus, S.G. Kovalenko, 
Phys. Rev. {\bf D 57} (1998) 1947

\bibitem[HM94]{101}
HEIDELBERG--MOSCOW collab., Phys. Lett. {\bf B 336} (1994) 141

\bibitem[HM95]{79}
HEIDELBERG--MOSCOW collab., Phys. Lett. {\bf B 356} (1995) 450

\bibitem[HM96]{HM96}
HEIDELBERG--MOSCOW collab., Phys. Rev. {\bf D 54} (1996) 3641, 

\bibitem[HM97]{HM97}
HEIDELBERG--MOSCOW collab., Phys. Rev. {\bf D 55} (1997) 54 and Phys. Lett.
{\bf B 407} (1997) 219 

\bibitem[HM98]{HM98}
HEIDELBERG--MOSCOW collab., Phys. Rev. {\bf D 59} (1998) 022001-1  

\bibitem[Hug60]{rel}  V.W. Hughes, H.G. Robinson, and V.
  Beltran-Lopez, Phys. Rev. Lett. {\bf 4}, 342 (1960); R.W.P.
  Drever, Philos. Mag. {\bf 6}, 683 (1961); D. Newman, G.W. Ford, A. Rich
  and E. Sweetman, Phys. Rev. Lett. {\bf 40}, 1355 (1978); A. Brillet and J.L.
  Hall, Phys. Rev. Lett. {\bf 42}, 549 (1979); J.D. Prestage, J.J.
  Bollinger, W.M. Itano, and D.J. Wineland, Phys. Rev. Lett. {\bf
   54}, 2387 (1985); S.K. Lamoureaux, J.P. Jacobs, B.R. Heckel,
  R.J. Raab, and E.N. Fortson, Phys. Rev. Lett. {\bf 57}, 3125 (1986).



\bibitem[Ioa94]{21}
A. Ioanissyan, J.W.F. Valle, Phys. Lett {\bf B 322} (1994) 93

\bibitem[Joer94]{95} V. J\"orgens et al., Nucl. Phys. (Proc. Suppl.) {\bf B  35} (1994) 378 

\bibitem[Jun96]{Jun96}
G. Jungmann, M. Kamionkowski, K. Griest, Phys. Rep. 267 (1996) 195


\bibitem[Kal97]{Kal97}
J. Kalinowski et al., preprint hep-ph/9703288v2 (March 1997)

\bibitem[Kan97]{Kan97}
G. Kane, in \cite{Kla98}


\bibitem[Kla87]{84}
H.V. Klapdor--Kleingrothaus, MPI--H 1987, proposal

\bibitem[Kla91]{Kla91}
H.V. Klapdor--Kleingrothaus, Proc. Int. Symposium on $\gamma$-Ray Astrophysics,
Paris 1990, AIP Conf. Proc. 232 (1991) 464

\bibitem[Kla92]{Kla92}
H.V. Klapdor--Kleingrothaus, K. Zuber, Phys. Bl. {\bf 48} (1992) 1017 

\bibitem[Kla94]{81}
H.V. Klapdor--Kleingrothaus, Progr. Part. Nucl. Phys. {\bf 32} (1994)261


\bibitem[Kla95]{1}
H. V. Klapdor--Kleingrothaus, A. Staudt, Non--Accelerator Particle Physics,
IOP Publ., Bristol, Philadelphia, 1995; and Teilchenphysik ohne Beschleuniger,
Teubner Verlag, Stuttgart, 1995

\bibitem[Kla96a]{tren}
H.V. Klapdor--Kleingrothaus and S. Stoica (Eds.), Proc. {\it Int. Workshop
on Double Beta Decay and Related Topics}, Trento, 24.4.--5.5.95, 
World Scientific Singapore


\bibitem[Kla96b]{Kla95b}
H.V. Klapdor--Kleingrothaus, in Proc. 
{\it Int. Workshop
on Double Beta Decay and Related Topics}, Trento, 24.4.--5.5.95, 
World Scientific Singapore 1996, Ed.: H.V. Klapdor--Kleingrothaus and S. Stoica


\bibitem[Kla97a]{Kla97}
H.V. Klapdor--Kleingrothaus, Invited talk at NEUTRINO 96, Helsinki, 
June 1996, World Scientific Singapore 1997, p. 317


\bibitem[Kla97b]{Kla97b}
H.V. Klapdor--Kleingrothaus, M.I. Kudravtsev, V.G. Stolpovski,
S.I. Svertilov, V.F. Melnikov, I. Krivosheina, J. Moscow. Phys. Soc.
7 (1997) 41


\bibitem[Kla97c]{Kla97d}
H.V. Klapdor--Kleingrothaus, M. Hirsch, Z. Phys. {\bf A 359} (1997) 361 

\bibitem[Kla97d]{Kla97e}
H.V. Klapdor--Kleingrothaus, Y. Ramachers, in: Proc. 
Int. Workshop on
Dark Matter in Astro-- and Particle Physics (DARK96) Sept. 1996,
Eds. H.V. Klapdor--Kleingrothaus and Y. Ramachers, Heidelberg, 
(World Scientific Singapore) 1997, p. 459

\bibitem[Kla98]{Kla98}
H.V. Klapdor--Kleingrothaus and H. Paes (Eds.), Beyond the Desert --
Accelerator- and Non--Accelerator Approaches, IOP, Bristol, 1998

\bibitem[Kla98a]{KK1}
H.V. Klapdor--Kleingrothaus, in \cite{Kla98}

\bibitem[Kla98b]{KK2}
H.V. Klapdor--Kleingrothaus, Int. J. Mod. Phys. {\bf A 13} (1998) 3953-3992

\bibitem[Kla98c]{KK3}
H.V. Klapdor--Kleingrothaus, J. Hellmig, M. Hirsch,
J. Phys. {\bf G 24} (1998) 483-516

\bibitem[Kla98d]{Kla98d}
H.V. Klapdor--Kleingrothaus, Y. Ramachers, Eur. Phys. J. {\bf A 3} (1998) 85

\bibitem[Kla98e]{KKP}
H.V. Klapdor--Kleingrothaus, H. P\"as;
in: Proc. of the 6th Symp. on Particles, Strings and Cosmology (PASCOS'98), 
Boston/USA, 1998

\bibitem[Kla98f]{KPS}
H.V. Klapdor--Kleingrothaus, H. P\"as, U. Sarkar, Eur. Phys. J. {\bf A 5} 
(1999) 3, and hep-ph/9809396

\bibitem[Kla98g]{klapneut}
H.V. Klapdor--Kleingrothaus, Proc. NEUTRINO'98 \\(Takayama,Japan,June 1998)
World Scientific, Singapore (1999)

\bibitem[Kla98h]{Kla98h}
H.V. Klapdor--Kleingrothaus et al., Adv. Space Res. {\bf 21} (1998) 347

\bibitem[Kla99a]{Kla99a}
H.V. Klapdor--Kleingrothaus, Proc. LEPTON and BARYON NUMBER VIOLATION in 
Particle Physics, Astrophysics and Cosmology, Trento, Italy, 20-25 April 1998, 
eds. H.V.Klapdor-Kleingrothaus and I.V.Krivosheina, IOP, (1999) 251

\bibitem[Kla99b]{Kla99b}
H.V. Klapdor--Kleingrothaus, H. P\"as, A.Yu. Smirnov, to be publ. 


\bibitem[Kol97]{Kol97a}
S. Kolb, M. Hirsch, H.V. Klapdor--Kleingrothaus, Phys. Rev. {\bf D 56} (1997) 
4161

\bibitem[Kuc95]{Kuc95}
R. Kuchimanchi, R.N. Mohapatra, Phys. Rev. Lett. {\bf 75} (1995) 3939



\bibitem[Kum96]{96} K. Kume (ELEGANT collaboration), in \cite{tren}


\bibitem[Kuz90]{18}
V. Kuzmin, V. Rubakov, M. Shaposhnikov, Phys. Lett. {\bf B 185} (1985)
36; M. Fukugita, T. Yanagida, Phys. Rev. {\bf D 42} (1990) 1285;
G. Gelmini, T. Yanagida, Phys. Lett. {\bf B 294} (1992) 53; 
B. Campbell {\it et al.}, Phys. Lett. {\bf B 256} (91) 457



\bibitem[Lan88]{15}
P.Langacker, in `Neutrinos' (Springer: Heidelberg, New York), 1988,
ed. H.V. Klapdor, p. 71

\bibitem[Lee94]{19}
D.G. Lee, R.N. Mohapatra, Phys. Lett. {\bf B 329} (1994) 463


\bibitem[Ma99]{Ma99}
E. Ma, hep-ph/9902392

\bibitem[Min97]{Min97}
H. Minakata, O. Yasuda, hep-ph/9712291

\bibitem[Moe91]{99} M. K. Moe, Phys. Rev. {\bf C 44} (1991) R931 

\bibitem[Moe94]{93} M. K. Moe, Prog. Part. Nucl. Phys. {\bf 32} (1994) 247; 
Nucl. Phys. (Proc. Suppl.) {\bf B 38} (1995) 36


\bibitem[Moh86a]{48}
R.N. Mohapatra, Phys. Rev. {\bf D 34} (1986) 3457

\bibitem[Moh86b]{50}
R.N. Mohapatra, Phys. Rev. {\bf D 34} (1986) 909 


\bibitem[Moh88]{73}
R.N. Mohapatra and E. Takasugi, Phys. Lett. {\bf B 211} (1988) 192


\bibitem[Moh91]{17}
R.N. Mohapatra, P.B. Pal, 
'Massive Neutrinos in Physics and Astrophysics'
(World Scientific, Singapore) 1991

\bibitem[Moh92]{45}
R.N. Mohapatra, Unification and Supersymmetry (Springer: Heidelberg, New York)
1986 and  1992

\bibitem[Moh94]{Moh94}
R.N. Mohapatra, Progr. Part. Nucl. Phys. {\bf 32} (1994) 187

\bibitem[Moh95]{23}
R.N. Mohapatra, S. Nussinov, Phys. Lett. {\bf B 346} (1995) 75

\bibitem[Moh96a]{Moh96}
R.N. Mohapatra, A. Rasin, Phys. Rev. Lett. {\bf 76} (1996) 3490 and Phys. Rev. 
{\bf D54} (1996) 5835

\bibitem[Moh96b]{7}
R.N. Mohapatra, in \cite{tren}


\bibitem[Moh97a]{Mohneu}
R. N. Mohapatra, Proc. {\it Neutrino 96}, Helsinki, 1996, World 
Scientific Singapore, 1997, p. 290

\bibitem[Moh97b]{Moh97a}
R.N. Mohapatra, Proc. Int. School on Neutrinos, 
Erice, Italy, Sept. 1997, Progr. Part. Nuc. Phys. {\bf 40} (1998)


\bibitem[Mut88]{27}
K. Muto, H.V. Klapdor, in 'Neutrinos (Springer: Heidelberg, 
New York) 1988, ed. H.V. Klapdor, p. 183

\bibitem[Mut89]{28}
K. Muto, E. Bender, H.V. Klapdor, Z. Phys. {\bf A 334} (1989) 177,187;


\bibitem[NEM94]{94} NEMO Collaboration, Nucl. Phys. (Proc. Suppl.) {\bf B 35} 
(1994) 369 

\bibitem[Nar95]{Nar95} 
E. Nardi {\it et al.}, Phys. Lett. {\bf B 344}(1995) 225 

\bibitem[Nor97]{Nor97}
D. Normile, Science 276 (1997) 1795 

\bibitem[PDG94]{46}
Particle Data Group, Phys. Rev. {\bf D 50} (1994)


\bibitem[P\"as96]{71}
H. Paes {\it et al.}, in \cite{tren};

\bibitem[P\"as97]{Paes97}
H. Paes, M. Hirsch, H.V. Klapdor--Kleingrothaus, S.G. Kovalenko,
in \cite{Kla98}

\bibitem[P\"as99]{Paes99}
H. P\"as, M. Hirsch, H.V. Klapdor--Kleingrothaus, S.G. Kovalenko,
Phys. Lett. {\bf B 453} (1999) 194 

\bibitem[Pan96]{8}
O.Panella, in \cite{tren}

\bibitem[Pan97a]{Pan97}
O. Panella, in \cite{Kla98}

\bibitem[Pan96b]{Pan96}
G. Pantis, F. Simkovic, J.D. Vergados, A. Faessler, Phys. Rev. {\bf C 53} 
(1996) 695

\bibitem[Pan99]{Pan99}
O. Panella {\it et al.}, hep-ph/9903253v2

\bibitem[PDG98]{PDG98}
Particle Data Group, Eur. Phys. J. {\bf C 3} (1998) 1

\bibitem[Pel93]{Pel93}
J. Peltoniemi, J. Valle, Nucl. Phys. {\bf B 406} (1993) 409

\bibitem[Pel95]{76}
J.T. Peltoniemi, preprint hep-ph/9506228


\bibitem[Pet94]{20}
S.T. Petcov, A.Yu. Smirnov, Phys. Lett. {\bf B 322} (1994) 109



\bibitem[Pet96]{12}
S. Petcov, in \cite{tren}



\bibitem[Piq96]{98} F. Piquemal {\it et al.}, in \cite{tren}  



\bibitem[Pri98]{Pri98}
J.R. Primack, M.A.K. Gross, astro-ph/9810204






\bibitem[Rag94]{97} R. S. Raghavan, Phys. Rev. Lett. {\bf 72} (1994) 1411 








\bibitem[Riz96]{Riz96}
T.G. Rizzo, hep/ph/9612440



\bibitem[Sch81]{Sch81}
J. Schechter, J.W.F. Valle, Phys. Rev. {\bf D 25} (1982) 2951






\bibitem[Smi96]{Smi96a}
A. Yu. Smirnov, Proc. Int. Conf. on High Energy Physics, 
Warsaw 1996, hep-ph/9611465v2
(Dec 1996)

\bibitem[Sou92]{51}
I.A. D'Souza. C.S. Kalman, Preons, Models of Leptons , 
Quarks and Gauge bosons
as Composite Objects (World Scientific, Singapore) 1992



\bibitem[Sta90]{29}
A. Staudt, K.Muto, H.V. Klapdor--Kleingrothaus, 
Europhys. Lett. {\bf 13} (1990) 31










\bibitem[Suz97]{Suz97}
A. Suzuki, priv. comm. 1997, and KAMLAND proposal, (in Japanese)


\bibitem[Tak96]{9}
E.Takasugi, in \cite{tren}  

\bibitem[Tak97]{Tak97}
E. Takasugi, in \cite{Kla98}



\bibitem[Tre95]{83}
V.I. Tretyak, Yu. Zdesenko, At. Data Nucl. Data Tables {\bf 61} (1995) 43


\bibitem[Val96]{13}
J.W.F. Valle, in \cite{tren}

\bibitem[Vis99]{Vis99}
F. Vissani, hep-ph/9906525




\bibitem[Vui93]{92} J.-C. Vuilleumier et al., Phys. Rev. {\bf D 48} (1993) 1009

\bibitem[Wil92]{will} C. M. Will, {\it Theory and Experiment in Gravitational 
  Physics}, 2nd edition (Cambridge University Press, Cambridge, 1992).



\bibitem[Wol81]{Wol81} L. Wolfenstein, Phys. Lett {\bf 107 B} (1981) 77

\bibitem[You95]{87}
Ke You et al., Phys. Lett. {\bf B 265} (1995) 53 


\end{thebibliography}
\end{document}